\documentclass[reprint,superscriptaddress,amsmath,amssymb]{revtex4-2}

\usepackage{graphicx}
\usepackage{dcolumn}
\usepackage{bm}
\usepackage[driverfallback=dvipdfmx,colorlinks,linkcolor=blue,citecolor=blue,urlcolor=blue]{hyperref}
\usepackage{ulem}
\usepackage[english]{babel}
\usepackage{svg}
\usepackage[final]{changes}

\begin{document}

\title{Second-harmonic generation from an optically levitated KTP nanocrystal in vacuum}

\author{Yuanbin Jin}
\email{jinyb@sxu.edu.cn}
\thanks{These authors contributed equally to this work.}
\affiliation{State Key Laboratory of Quantum Optics Technologies and Devices, Institute of Opto-Electronics, Shanxi University, Taiyuan, Shanxi 030006, China}

\author{Chenli Gao}
\thanks{These authors contributed equally to this work.}
\affiliation{State Key Laboratory of Quantum Optics Technologies and Devices, Institute of Opto-Electronics, Shanxi University, Taiyuan, Shanxi 030006, China}

\author{Jiayu Feng}
\thanks{These authors contributed equally to this work.}
\affiliation{State Key Laboratory of Quantum Optics Technologies and Devices, Institute of Opto-Electronics, Shanxi University, Taiyuan, Shanxi 030006, China}

\author{Yuxuan Hao}
\affiliation{State Key Laboratory of Quantum Optics Technologies and Devices, Institute of Opto-Electronics, Shanxi University, Taiyuan, Shanxi 030006, China}

\author{Xudong Yu}
\email{jiance\_yu@sxu.edu.cn}
\affiliation{State Key Laboratory of Quantum Optics Technologies and Devices, Institute of Opto-Electronics, Shanxi University, Taiyuan, Shanxi 030006, China}

\author{Jing Zhang}
\email{jzhang74@sxu.edu.cn}
\affiliation{State Key Laboratory of Quantum Optics Technologies and Devices, Institute of Opto-Electronics, Shanxi University, Taiyuan, Shanxi 030006, China}
\affiliation{Hefei National Laboratory, Hefei, Anhui 230088, China}


\begin{abstract}
The optically levitated system in vacuum has emerged as a powerful platform for studies of fundamental physics and precision measurements. Although various nanoparticles have been successfully levitated in vacuum, they typically lack the capability to support optical nonlinear processes. Here, we experimentally demonstrate the stable levitation of a potassium titanyl phosphate (KTP) nonlinear nanocrystal in vacuum and investigate its second-harmonic generation (SHG) properties. This levitated system intrinsically provides a pristine dark-background environment with a high signal-to-noise ratio. The trapping laser simultaneously serves as a fundamental light for efficient SHG. Moreover, the polarization of the collected SHG signal is correlated with that of the fundamental laser, providing clear evidence of the optical torque enabling controllable alignment of the nanocrystal with the driving field. Our work establishes a new route toward exploring nonlinear optical processes in vacuum levitation systems and designing novel nanodevices with high manipulation agility in a fully contact-free environment.
\end{abstract}

\maketitle

\section{Introduction}

Vacuum optical levitation employs tightly focused laser beams to trap, suspend, and manipulate microscopic objects. It offers exceptional isolation from environmental noise and precise control over decoherence induced by background gas collisions \cite{GonzalezLevitodynamics2021,VolpeRoadmap2023,JinTowards2024}. The center-of-mass (CoM) motion of a levitated particle can be controlled with extraordinary precision, enabling ultrahigh quality factors and extreme sensitivities in force, torque and acceleration sensing \cite{LiMeasurement2010,MonteiroForce2020,AhnUltrasensitive2020,LiangYoctonewton2023}. Rapid advances have propelled the field toward new frontiers. By employing a variety of cooling techniques, the CoM motion of levitated particles has been cooled to the quantum ground state \cite{DelicCooling2020,MagriniReal2021,TebbenjohannsQuantum2021}, paving the way for the exploration of macroscopic quantum phenomena. Concurrently, optically levitated nanoparticles have been driven to rotate at GHz frequencies \cite{JinGHz2021,AhnOptically2018,AhnUltrasensitive2020}, enabling nanoscale gyroscopes and facilitating proposals for probing non-Newtonian gravity, Casimir torque, and vacuum friction \cite{GeraciShort2010,ChenConstraining2022,SomersMeasurement2018,XuEnhancement2021}.

The low absorption of silica at infrared trapping wavelengths makes it a widely adopted material for this system. Nevertheless, it faces difficulties, such as the absence of manipulating means of the internal degree of freedom and the potential of more function extension application for this isotropic particle. Recently, some studies have extended to other materials for exploring much richer physics and questing new applications. The diamond hosting nitrogen-vacancy (NV) centers, rare-earth-doped $NaYF_{4}$, and $YLF$ particles have been stably levitated \cite{JinQuantum2024,NeukirchMulti2015,HoangElectron2016,WinstoneOptical2022,RahmanLaser2017}, presenting unique ways for studying the coupling of internal and external degrees of freedom, geometric phases, internal temperature cooling, quantum rotor, and so on \cite{DelordSpin2020,MaclaurinMeasurable2012,RahmanLaser2017,SticklerQuantum2021}. Nonlinear optical materials represent a promising frontier \cite{YangThe2020,ChenHigh2021,ChenOptical2025,BehelControlled2024}. While bulk crystals such as KTP are well-established, extending them to the micro- and nanoscale enables integration into on-chip photonic platforms \cite{LePhotostable2008,ZhangThree2011,BonacinaHarmonic2020}. Meanwhile, the strong SHG from KTP nanocrystals deposited on a glass surface is, in particular, easily detectable and allows precise orientation mapping via nonlinear polarimetry and defocused imaging \cite{LePhotostable2008,SandeauDefocused2007}. In the previous studies, a persistent challenge at the nanoscale has been the degradation of optical properties due to particle-surface interactions, along with the difficulty in achieving dynamic control. To date, nonlinear optical processes in optically levitated systems have remained unexplored.

In this work, we merge the fields of levitated optomechanics and nonlinear optics at the nanoscale by optically trapping a KTP nanocrystal in vacuum\replaced{. Compared with substrate-supported nanocrystals, optical levitation eliminates particle–surface interactions and provides several advantages: (i) the rotational degrees of freedom remain accessible, enabling optical torque control of the crystal orientation; (ii) the optical field simultaneously traps the particle, controls its orientation, and serves as the fundamental field for SHG, enabling dynamic tuning of the nonlinear response; and (iii) the accessible mechanical degrees of freedom may provide a platform for studying the coupling between particle dynamics and nonlinear optical process.}{ , which completely eliminates the surface-interaction.} We demonstrate efficient SHG, where the SHG signal is collected and characterized with a single-photon counter and a spectrometer. The polarization of the second-harmonic emission aligns with that of the trapping laser, as well as the libration signal of the levitated nanocrystal-both mutually providing direct evidence of the torque that dynamically aligns the nanocrystal principle axis with the fundamental laser. The optically levitated nonlinear nanocrystals in vacuum demonstrates a novel and potential platform for studies of nanolasers and nanoscaled nonlinear optical devices, free from the constraints of a substrate.

\section{SHG from a levitated KTP nanocrystal}

KTP crystals are widely applied in nonlinear optics because of its broad transparency window (350–4500 nm), large nonlinear optical coefficients, and high damage threshold under intense optical fields. In this work, KTP nanocrystals are \replaced{prepared by mechanically grinding a bulk KTP crystal into fine particles, as described in detail in the Methods section,}{ fabricated by mechanically milling a bulk crystal} and subsequently optically levitated in vacuum using a strongly focused 1064 nm laser beam. The trapping laser simultaneously serves as both the optical trap and the fundamental light. The resulting SHG signal at 532 nm, arising from a two-photon nonlinear process, is collected and characterized. The levitated nanocrystals have characteristic dimensions on the order of 100 nm, which depends on the numerical aperture (NA) of the objective lens and the trapping laser power. This is consistent with previous studies on levitated silica nanoparticles \cite{Jinoptically2018,TzarouchisLight2018}. For particles at this scale, the conventional phase-matching condition required in bulk nonlinear crystals can be neglected due to the sub-wavelength particle dimensions relative to the fundamental laser wavelength. \replaced{In this regime}{ Consequently}, the SHG emission intensity distribution exhibits a radiation pattern analogous to that of a dipole scatter \added{\cite{LePhotostable2008,SandeauDefocused2007}}. \added{The reverse process of SHG, parametric down-conversion \cite{WuGeneration1986,YuExperimental2014}, may offer a promising route toward developing levitated nonlinear nanocrystals as sources of nonclassical light.}

KTP is a non-centrosymmetric orthorhombic crystal. Under an incident optical field ${\bf{E}}\left( {{E_X},{E_Y},{E_Z}} \right)$ at the fundamental frequency $\omega$, the crystal behaves as a second-order nonlinear dipole source. The induced second-order polarization $P_i^{(2)}$ is expressed as
\begin{equation}
	P_i^{\left( 2 \right)} = {\varepsilon _0}\sum\limits_{j,k} {\chi _{ijk}^{\left( 2 \right)}{E_j}{E_k}}
	\label{eq1},
\end{equation}
where $\varepsilon_0$ is the vacuum permittivity and $\chi_{ijk}^{(2)}$ denotes the second-order susceptibility tensor, with indices $i$, $j$, and $k$ corresponding to the crystal axes ($X$, $Y$, and $Z$). This crystal coordinate system differs from the laboratory frame defined by ($x$, $y$, and $z$). Among the tensor elements, $\chi_{ZZZ}$ is the largest nonlinear coefficient of KTP crystal.

\begin{figure*}[t]
	\centering\includegraphics[width=15cm]{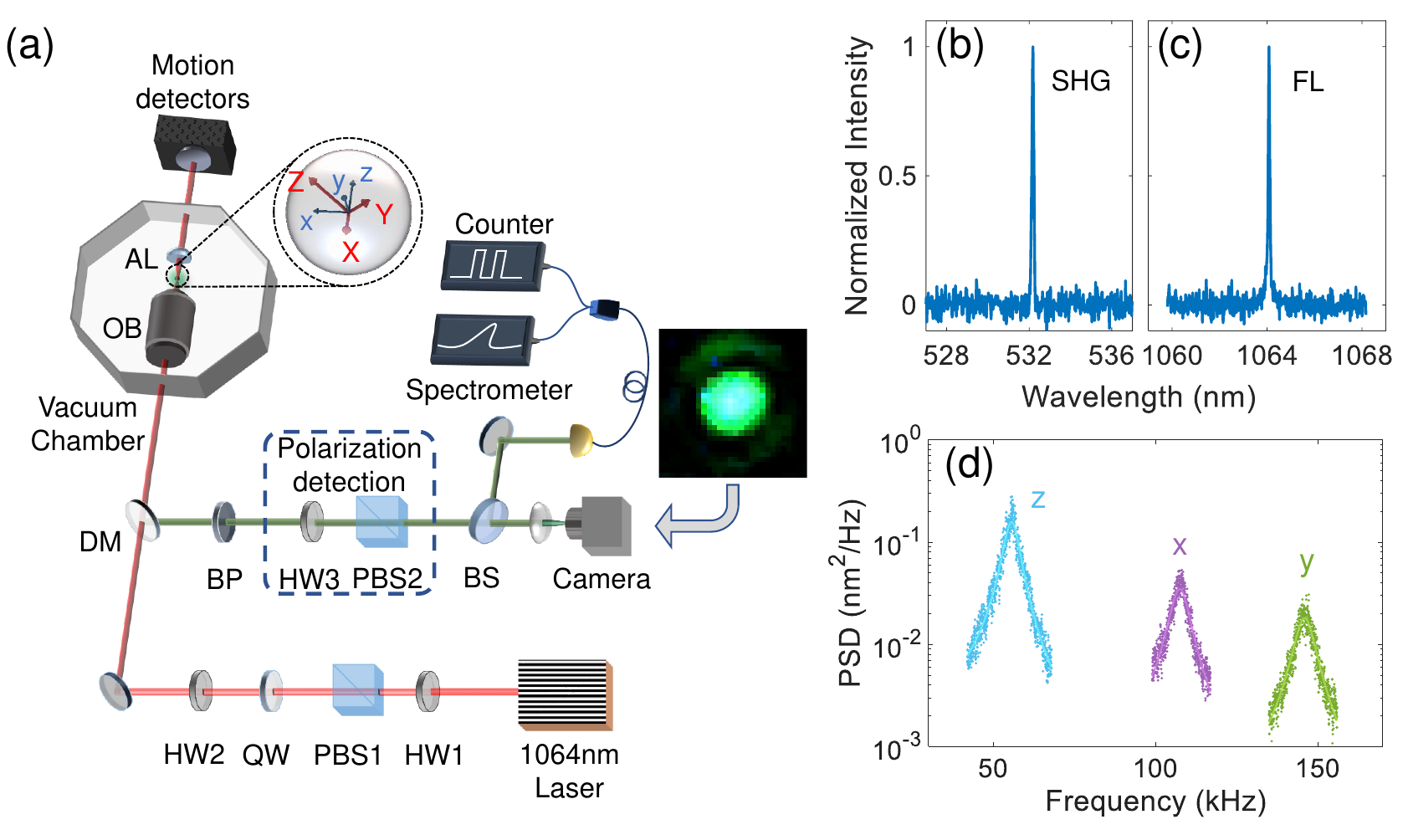}
	\caption{\label{fig:setup} Optical levitation and second-harmonic generation (SHG) from a KTP nanocrystal in vacuum. (a) Schematic of the experimental setup. A 1064 nm laser with tunable power and polarization is focused by an objective lens (NA = 0.95) to optically levitate a KTP nanocrystal in vacuum. The forward-scattered light is collected by an aspheric lens (AL) and detected using a balanced homodyne system to monitor the particle motion. The backward-propagating SHG signal is collected by the objective lens and directed to a camera, a single-photon counter, and a spectrometer for imaging, intensity measurements, and spectral characterization, respectively. The inset shows a camera image of the SHG signal from a levitated KTP nanocrystal. DM, dichroic mirror; HW1–HW3, half-wave plates; QW, quarter-wave plate; BP, bandpass filter; BS, beam splitter. Two coordinate systems are defined: the laboratory frame ($x,y,z$) and the crystal frame ($X,Y,Z$). (b) Optical spectrum of the SHG signal centered at 532 nm. (c) Optical spectrum of the 1064 nm fundamental laser (FL). (d) Power spectral densities (PSDs) of the center-of-mass (CoM) motion of the levitated KTP nanocrystal in the $x$, $y$, and $z$ directions at a pressure of 1 kPa.}
\end{figure*}

KTP is a biaxial crystal whose largest refractive index lies along the $Z$ axis. The refractive index anisotropy gives rise to an alignment torque that tends to orient this axis parallel to the polarization of the fundamental field. Therefore, we assume that the major axis of the polarization ellipse coincides with the crystal $Z$ axis, such that the laser propagates in the $XY$ plane and forms an angle $\varphi$ with the $X$ axis. Further derivation of Eq. (\ref{eq1}) leads to
\begin{equation}
	{P^{\left( 2 \right)}} = 2{\varepsilon _0}E_0^2{e^{ - 2i\omega t}}\left[ {{A_Z}{{\bf{e}}_Z} \pm i\left( {{A_X}{{\bf{e}}_X} + {A_Y}{{\bf{e}}_Y}} \right)} \right]
	\label{eq2},
\end{equation}
where
\begin{equation}
	{A_Z} = \left[ {{d_{33}}{{\cos }^2}\theta  - \left( {{d_{31}}{{\cos }^2}\varphi  + {d_{32}}{{\sin }^2}\varphi } \right){{\sin }^2}\theta } \right]
	\label{eq3},
\end{equation}
\begin{equation}
	{A_X} = 2{d_{15}}\cos \varphi \sin \theta \cos \theta
	\label{eq4},
\end{equation}
\begin{equation}
	{A_Y} = 2{d_{24}}\sin \varphi \sin \theta \cos \theta
	\label{eq5},
\end{equation}
where $E_0$ is the amplitude of the fundamental electric field. The second-order susceptibility tensor is related to the reduced matrix through ${d_{il}} = \chi _ {ijk}^{\left( 2 \right)}/2$, following the convention summarized in Table S1. Here, $\cos \theta = {E_{0Z}}/{E_0}$ characterizes the ellipticity of the incident laser polarization, with $E_{0Z}$ representing the $Z$ component of the electric field. The unit vectors along the crystal axes are denoted by ${\bf{e}}_X$, ${\bf{e}}_Y$, and ${\bf{e}}_Z$, respectively. The positive and negative signs correspond to left- and right-handed polarization states. Therefore, the SHG signal can be modeled as a dipole emitter whose polarization axis is determined by $P^{\left( 2 \right)}$, giving an intensity proportional to the square of $P^{\left( 2 \right)}$,
\begin{equation}
	{I^{\left( 2 \right)}} \propto {\left| {{P^{\left( 2 \right)}}} \right|^2} \propto {\left( {{I^{\left( 1 \right)}}} \right)^2}
	\label{eq6}.
\end{equation}
Accordingly, the SHG intensity directly depends on both the power and polarization state of the fundamental laser. Moreover, Eqs. (\ref{eq2})–(\ref{eq5}) show that the polarization of the SHG emission is determined by that of the fundamental laser. Under a linearly polarized fundamental laser, the SHG emerges linearly polarized along the same direction. For an elliptically polarized fundamental laser, the SHG remains elliptically polarized, with its ellipticity determined by the relative amplitudes of the polarization components of $P^{\left( 2 \right)}$.

Figure \ref{fig:setup}(a) is the schematic of the experimental setup. The 1064 nm beam from a solid-state laser first passes through a half-wave plate (HW1) and a polarizing beam splitter (PBS1) for power control. A quarter-wave plate (QW) and a second half-wave plate (HW2) are then used to define the polarization state of the trapping beam. The beam is tightly focused by a high-NA objective lens (NA = 0.95) to form the optical trap. The forward fundamental light is collected by an aspheric lens and directed to a balanced homodyne detection system for monitoring the CoM motions and librational motions of the levitated nanocrystal (Fig. S1) \cite{JinGHz2021}. The backward-propagating SHG signal is collected by the same objective lens. A bandpass filter (BP) centered at 532 nm with a full width at half maximum (FWHM) bandwidth of 10 nm is employed to suppress the residual 1064 nm trapping laser. The filtered SHG signal is subsequently routed to a camera for real-time imaging, a single-photon counter for intensity measurements, and a spectrometer for spectral characterization.

The KTP nanocrystals are prepared by mechanically milling a high-quality bulk crystal. The resulting nanoparticles are dispersed in ethanol and introduced into the vacuum chamber using an ultrasonic nebulizer. Successful trapping of a KTP nanocrystal is initially identified by the appearance of a stable green spot on a color camera, as shown in the inset of Fig. \ref{fig:setup}(a). Since the bandpass filter blocks the scattered 1064 nm laser, the observed green emission corresponds to the SHG signal. To further verify the presence of a KTP nanocrystal, the fluorescence is analyzed using a spectrometer. As shown in Fig. \ref{fig:setup}(b), the measured optical spectrum exhibits a sharp peak at 532 nm, confirming that the signal is the second harmonic of the 1064 nm fundamental laser. The measured SHG linewidth is approximately 0.1 nm, limited by the resolution of the spectrometer. This is supported by the independently measured spectrum of the trapping laser shown in Fig. \ref{fig:setup}(c), which exhibits the same linewidth despite the intrinsic linewidth of the laser being only about 100 kHz. After spectral characterization, the SHG intensity is measured using a single-photon counter.

Figure \ref{fig:setup}(d) shows the power spectral densities (PSDs) of the CoM motion along the $x$, $y$, and $z$ directions. Here, the $y$ is defined opposite to the direction of gravity and the $z$ coincides with the propagation direction of the trapping laser. The trapping laser power is 103 mW, and the chamber pressure is approximately 1 kPa. In the experiment, a large number of KTP nanocrystals are characterized. Three representative particles are selected for detailed analysis and are discussed throughout this work. For particle 1, the measured mechanical oscillation frequencies are $\omega_x =2\pi {\times} 108$ kHz, $\omega_y = 2\pi {\times}146$ kHz, and $\omega_z = 2\pi {\times}56$ kHz. The particle dimensions can be calculated based on the measured damping rates along the three directions \cite{Jinoptically2018}. The resulting sizes along the $x$, $y$ and $z$ directions are approximately $r_x = 58.5$ nm, $r_y = 55.8$ nm, and $r_z = 70.6$ nm. These values confirm that the levitated KTP nanocrystal is substantially smaller than the fundamental laser wavelength (1064 nm), justifying the neglect of phase-matching condition in the nonlinear optical process.

\begin{figure*}[t]
	\centering\includegraphics[width=15cm]{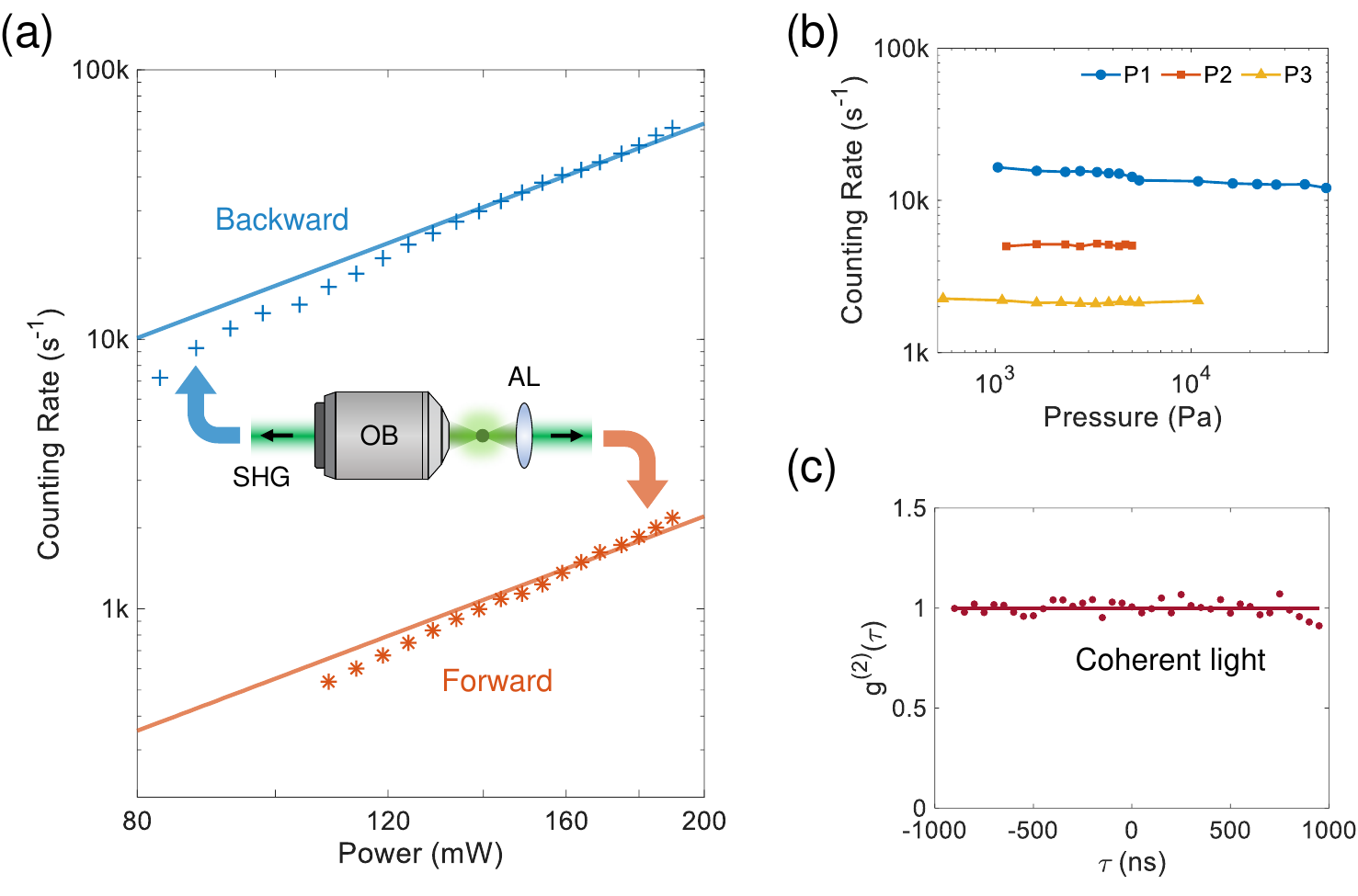}
	\caption{\label{fig:shg} SHG \replaced{characteristics}{ efficiency} and thermal stability of levitated KTP nanocrystals. (a) SHG photon counting rate as a function of the fundamental laser power at a pressure of 5 kPa. Blue crosses and red stars correspond to the backward- and forward-collected SHG signals. The solid curves are quadratic fits, confirming the second-order nonlinear nature of the SHG. (b) SHG photon counting rate \added{of the backward-collected signal} as a function of pressure at a laser power of 103 mW. The blue, red, and orange curves correspond to three randomly selected levitated KTP nanocrystals, each with dimensions of approximately 100 nm. The pressure independent SHG intensity demonstrates the excellent stability of the nonlinear optical response over a wide pressure range. (c) Second-order intensity-correlation function $g^{(2)}(\tau)$ of the SHG emission. This result confirms the coherent nature of the SHG emission.}
\end{figure*}

We investigate the key characteristics of the SHG emission, including its power dependence, polarization behavior, and operational stability. Figure \ref{fig:shg}(a) shows the SHG counting rate as a function of the fundamental laser power. The blue crosses represent the backward-collected SHG signal, which increases with laser power over the range from 80 mW to 190 mW. The solid fitting curve exhibits a clear quadratic dependence, confirming the characteristic second-order nonlinearity of the SHG process. \added{After accounting for the collection and detection efficiencies of the backward optical path, the total SHG photon generation rate is estimated to be approximately $4.5 \times 10^6$/s at a fundamental power of 120 mW, corresponding to an SHG conversion efficiency of $1.4 \times 10^{-11}$ (details are provided in Supplemental document).} The laser power cannot be reduced below a certain value in this configuration because sufficient optical power is required to maintain stable levitation. We also measure the forward-propagating SHG signal collected by the aspheric lens, shown as red stars in Fig. \ref{fig:shg}(a). The weaker forward signal mainly results from the smaller NA and lower collection efficiency of the aspheric lens. Nevertheless, the similar polarization dependence observed in both directions, together with the measured intensity ratio, supports a dipole-like radiation pattern for the SHG emission.

To evaluate the operational stability of the system, we investigate the SHG  response \added{of the backward-collected signal} as a function of chamber pressure. Figure \ref{fig:shg}(b) presents measurements obtained from three individual KTP nanocrystals with sizes around 100 nm. The SHG counting rate remains independent of pressure, demonstrating excellent stability of the nonlinear optical output. \added{The differences in SHG count rate among the three particles are attributed to particle-to-particle variations in volume and surface condition introduced during mechanical grinding.} As the pressure decreases, reduced gas-collision cooling combined with continuous optical absorption leads to \replaced{increased heating of the levitated nanocrystal. The particles become unstable and are eventually lost at pressures below 50 Pa, indicating significant laser induced heating at low pressures \cite{JinQuantum2024}. Nevertheless, the SHG response remains robust against the increased particle heating within the regime of stable optical trapping}{ an increase in particle temperature, until the particles become unstable and are lost at pressures near 50 Pa. Notably, this wide temperature variation does not compromise the stability of the SHG signal}. In addition, we measure the second-order correlation function \added{\cite{FishmanPhoton2023}} and obtain ${g^{\left( 2 \right)}}\left( \tau \right) = 1$, as shown in Fig. \ref{fig:shg}(c). \added{The detailed measurement is provided in the Supplementary Material.} This result confirms the coherent nature of the SHG emission.

\begin{figure*}
	\centering\includegraphics[width=14cm]{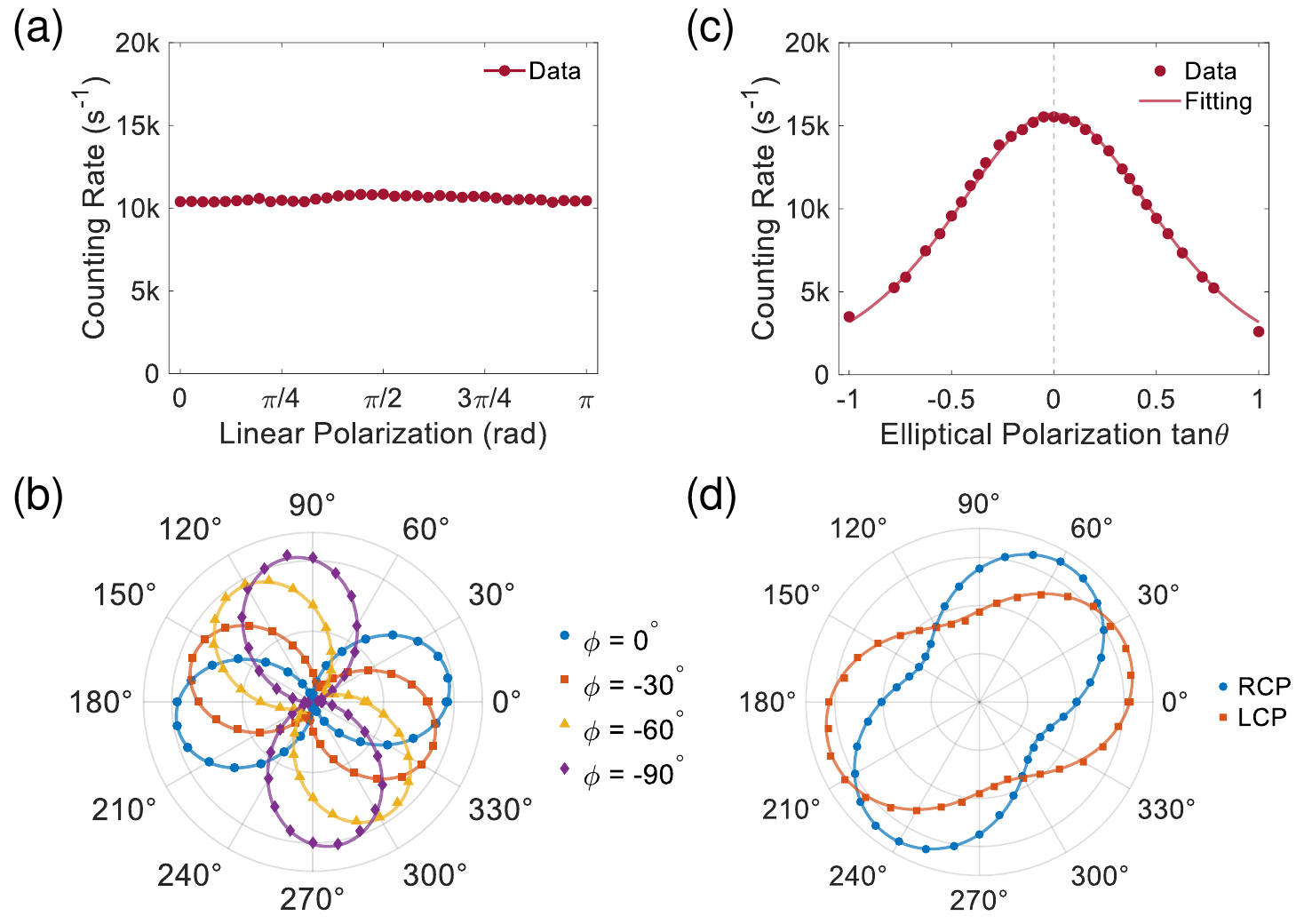}
	\caption{\label{fig:polarization} Polarization properties of SHG from a levitated KTP nanocrystal. (a) SHG intensity as a function of the linear polarization angle of the fundamental laser at a power of 85 mW. (b) Polarization-resolved SHG intensity in polar coordinates as a function of the analyzer angle, measured at the fundamental laser polarization angles $\phi$ of 0 (blue circles), $-\pi/6$ (red squares), $-\pi/3$ (orange triangles), and $-\pi/2$ (purple diamonds) relative to the $x$ axis. The solid curves are sinusoidal fits. (c) SHG intensity as a function of the fundamental polarization ellipticity at a power of 120 mW. The polarization of the fundamental laser is continuously varied from circular ($\tan \theta = -1 $) to linear ($\tan \theta = 0 $) and back to circular ($\tan \theta = +1 $). The solid curve is a fit based on Eq. (\ref{eq2}). (d) Polarization-resolved SHG intensity under right (blue circles) and left (red squares) circularly polarized fundamental laser, plotted in polar coordinates as a function of analyzer angle. The solid curves are sinusoidal fits.}
\end{figure*}

\section{Polarization characteristics of SHG}

A distinctive feature of SHG from levitated KTP nanocrystals, compared with their substrate-supported counterparts, is the ability to dynamically align the crystal optical axis with the polarization of the fundamental laser. To investigate this capability, we characterize the polarization properties of the SHG emission as functions of both the orientation and ellipticity of the fundamental laser polarization, as shown in Fig. \ref{fig:polarization}. We first examine the response to linearly polarized fundamental laser. By fixing the quarter-wave plate (QW) and rotating the half-wave plate (HW2), the linear polarization direction of the fundamental laser can be continuously varied. Without polarization-detection (in the absence of HW3 and PBS2), the total SHG intensity remains nearly unchanged as the polarization direction is rotated, as shown in Fig. \ref{fig:polarization}(a). This behavior provides the evidence that the levitated nanocrystal undergoes rotational reorientation, continuously aligning its optical axis with the polarization direction of the fundamental field. The minor intensity variation is attributed to slight change in polarization ellipticity introduced by mirror reflections and tight focusing through the objective lens.

To determine the polarization state of the SHG emission, a half-wave plate (HW3) and a polarizing beam splitter (PBS2) are inserted into the detection path. Figure \ref{fig:polarization}(b) presents the polarization-resolved SHG intensity in polar coordinates as a function of the analyzer angle for several orientations of the fundamental laser polarization (blue circles: 0, red squares: $-\pi/6$, orange triangles: $-\pi/3$, and purple diamonds: $-\pi/2$). The corresponding sinusoidal fits (solid curves) demonstrate that the SHG emission is linearly polarized and that its polarization direction closely follows that of the fundamental laser (Fig. S2). A small angular offset of approximately $9^\circ$ is observed, which we attribute to residual polarization distortions introduced by reflections from the optical components.

\begin{figure*}
	\centering\includegraphics[width=14cm]{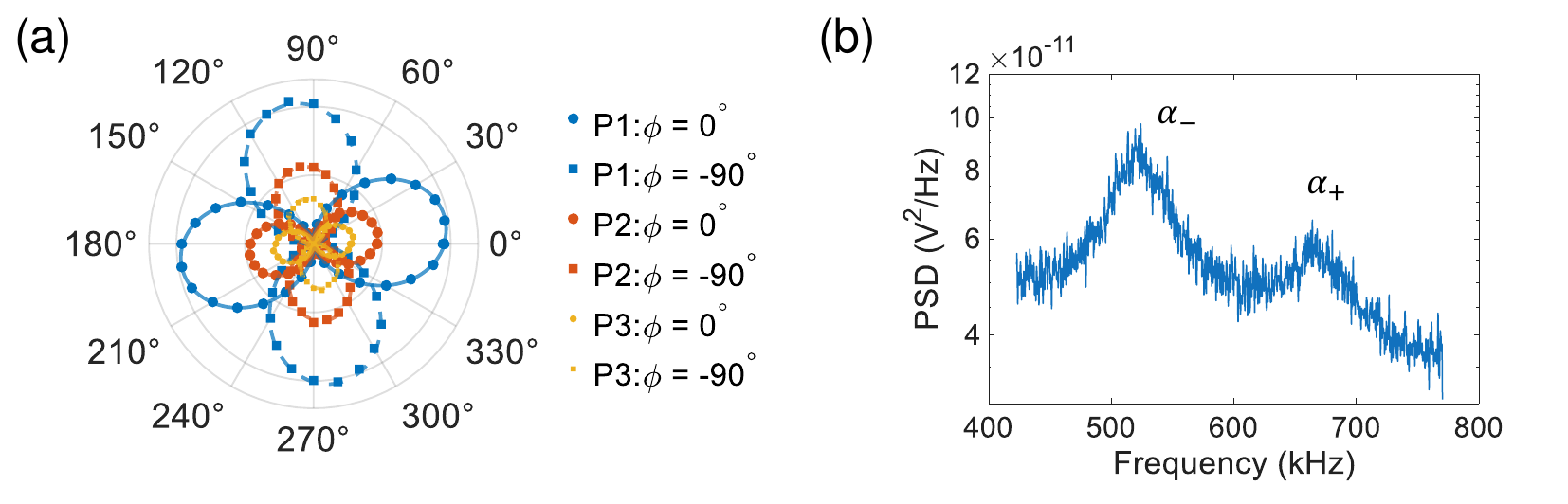}
	\caption{\label{fig:libration} Optical torque alignment of levitated KTP nanocrystals. (a) Polarization-resolved SHG intensity plotted in polar coordinates for the fundamental laser linear polarization angles $\phi$ of 0 (circles) and $-\pi/2$ (squares). Measurements from three individual levitated KTP nanocrystals are shown in blue, red, and orange. The solid curves are sinusoidal fits. \replaced{(b) }{(d)} Power spectrum of the librational motion along the $z$ axis of the laboratory frame with the linearly polarized fundamental laser in $x$ direction.}
\end{figure*}

We next investigate the influence of polarization ellipticity by adjusting the quarter-wave plate to continuously vary the fundamental laser polarization from circular to linear and back to circular. As shown in Fig. \ref{fig:polarization}(c), the measured SHG intensity strongly depends on the fundamental laser polarization state and is well described by Eq. (\ref{eq2}) (solid curve). The SHG intensity reaches its maximum under linear polarization and is approximately five times larger than that obtained under circular polarization. This pronounced dependence arises from the anisotropic nonlinear susceptibility of the KTP nanocrystal. \added{Under relatively high pressure conditions, the nanocrystal remains confined around a preferred orientation and undergoes librational motion rather than rotation.}

To further examine the polarization state of the SHG under circularly polarized excitation, polarization-resolved measurements are performed using HW3 and PBS2. Figure \ref{fig:polarization}(d) shows the measured intensity distributions for right-circularly polarized (blue circles) and left-circularly polarized (red squares) fundamental laser, together with the corresponding sinusoidal fits. \added{The polarization dependence and the well-defined polarization axis provide further evidence that the nanocrystal remains orientationally confined rather than undergoing rotation during the measurement.} In contrast to the linearly polarized SHG, the SHG generated by circularly polarized fundamental light exhibits an elliptical polarization state. This result is fully consistent with the nonlinear polarization model described by Eqs. (\ref{eq2})–(\ref{eq5}) and further confirms that the polarization properties of the SHG emission are governed by those of the fundamental field.

\section{Dynamic alignment of nonlinear nanocrystals}

The preceding results demonstrate that the optical axis of the levitated KTP nanocrystal continuously aligns with the polarization of the fundamental laser. Such alignment arises from polarization-dependent optical torque. The torque exerted by the optical field $\mathbf{E}$ on the nanocrystal can be written as
\begin{equation}
	{M} = \frac{1}{2}{\mathop{\rm Re}\nolimits} \left( {{\bf{p}}^* \times {\bf{E}}} \right)
	\label{eq7},
\end{equation}
where $\mathbf{p}$ is the induced dipole moment. For KTP nanocrystals, the induced dipole moment can be decomposed into contributions arising from shape anisotropy and refractive index anisotropy. Since the dipole moment associated with nonlinear optical processes is several orders of magnitude smaller than the linear induced dipole moment, its contribution to the optical torque is neglected. Thus,
\begin{equation}
	{\bf{p}} = {{\bf{p}}_s} + {\bf{p}}_r^{\left( 1 \right)}
	\label{eq8}.
\end{equation}
Details of the derivation are provided in the Supplemental document and the results are shown in Fig. S3.

To identify the dominant mechanism experimentally, we randomly characterized three levitated KTP nanocrystals with different shapes. Their polarization-resolved SHG responses are shown in Fig. \ref{fig:libration}(a), where the blue, red, and orange data points correspond to three individual particles, and circles and squares denote horizontal and vertical detection polarizations, respectively. Despite their different shapes\replaced{, as shown in the Methods section}{ (Table I)}, the corresponding sinusoidal fits (solid curves) shows that all three particles exhibit the same polarization alignment behavior (Fig. S4). In practice, the equilibrium orientation of a levitated KTP nanocrystal is determined by the combined action of both torque contributions. However, these observations indicate that the equilibrium orientation is governed predominantly by the refractive index anisotropy of KTP rather than by particle shape when the particle shape is nearly spherical. The $Z$ axis of the levitated nanocrystal dynamically aligns with the major axis of the polarization ellipse of the trapping laser, consistent with the theoretical model developed above.

Additionally, the power spectrum of the librational motion of particle 1 along the laboratory frame $z$ axis is shown in Fig. \ref{fig:libration}(b), measured under a linearly polarized fundamental laser with its polarization oriented along the $x$ direction. Two peaks are observed at 520 kHz ($\alpha_{-}$) and 668 kHz ($\alpha_{+}$), originating from the coupling of two librational modes in the presence of free rotation along the crystal $Z$ axis \cite{BangFive2020}. The corresponding libration frequency is approximately 589 kHz, indicating a strong restoring torque acting on the particle. Such a large libration frequency is consistent with the optical torque required to align the crystal $Z$ axis, which hosts the largest nonlinear coefficient of KTP, with the polarization of the fundamental field and thereby maximize the SHG efficiency. The splitting between the $\alpha_{-}$ and $\alpha_{+}$ modes reveals strong rotation-induced coupling of the librational degrees of freedom. This coupling encodes information about the otherwise difficult-to-access rotational dynamics around the crystal $Z$ axis and provides a potential route for investigating all six degrees of freedom in levitated systems \cite{PontinSimultaneous2023,TroyerQuantum2026}.

\section{Conclusion}

In conclusion, we demonstrate a novel platform based on optically levitated KTP nanocrystals in vacuum. Stable SHG at 532 nm is observed from a levitated nanocrystal. The emission preserves the narrow linewidth of the fundamental laser and exhibits remarkable stability across a wide pressure range. The contact-free vacuum environment eliminates particle-surface interactions, providing a pristine platform for investigating nonlinear optical processes at the nanoscale. By controlling the polarization state of the trapping laser, we achieve flexible manipulation and characterization of the nanocrystal orientation. The measured polarization properties of the SHG emission provide direct evidence that the optical alignment torque is dominated by the refractive index anisotropy of KTP rather than by particle shape. This torque-induced self-alignment enables dynamic matching between the crystal nonlinear optical axis and the fundamental field, thereby optimizing the nonlinear optical response.

Our results establish a direct link between levitated optomechanics and nonlinear optical processes. The combination of \replaced{a contact-free environment, dynamically accessible rotational degrees of freedom, and polarization-controlled nonlinear response distinguishes the levitated platform from substrate-supported nonlinear nanocrystals and highlights its potential as an isolated and dynamically tunable nonlinear nanoemitter}{  coherent nonlinear emission, polarization controllability, rotational manipulation, and a fully contact-free environment highlights the unique capabilities of levitated nonlinear nanocrystals as isolated nanoemitters}. This platform offers new opportunities for studying nonlinear light-matter interactions, nanoscale frequency conversion, and rotational optomechanics under well-controlled conditions. Furthermore, it provides a promising route toward the realization of highly coherent nanodevices, quantum light sources, and hybrid optomechanical systems for applications in precision measurement and quantum information.

\section{Methods}

The KTP nanocrystals used in this work are prepared by mechanical milling of a bulk KTP crystal. A small piece of high-quality KTP crystal is first crushed and ground using an agate mortar and pestle. To improve grinding efficiency and reduce particle loss, a small amount of ethanol is added during the milling process, which continues for approximately 3 h to obtain a relatively homogeneous particle size. The slurry is then placed in a bottle and subjected to ultrasonic treatment to break up particle agglomerates. After sonication, the suspension is allowed to stand for several hours to enable sedimentation-assisted size selection. Larger particles gradually accumulated at the bottom, while smaller particles remained suspended in the upper layer. The upper suspension is carefully collected, while the remaining sediment undergo additional grinding and sonication. This procedure is repeated three times to enrich the fraction of smaller particles. The final suspension contained KTP nanocrystals with a broad size distribution ranging from several tens of nanometers to a few micrometers. 



Prior to the optical levitation experiments, the particles are dispersed in ethanol and introduced into the vacuum chamber using an ultrasonic nebulizer. The size of particles that can be stably trapped is determined by the trapping efficiency of the optical tweezers, which depends on the NA of the objective lens and the trapping laser power. The levitated KTP nanocrystals investigated in this work typically have characteristic dimensions on the order of 100 nm. The grinding and sample-preparation procedures inevitably introduce a small fraction of impurity particles. Under the experimental conditions used here, approximately half of the trapped particles exhibit detectable SHG signals. The appearance of SHG provides a reliable indicator for identifying levitated KTP nanocrystals.

The dimensions of the three representative levitated KTP nanocrystals studied in this work are \replaced{$r_x=58.5$ nm, $r_y=55.8$ nm, and $r_z=70.6$ nm for P1; $r_x=42.3$ nm, $r_y=34.7$ nm, and $r_z=35.4$ nm for P2; $r_x=56.6$ nm, $r_y=50.9$ nm, and $r_z=49.9$ nm for P3}{ summarized in Table I}. These values are estimated from the damping rates of the corresponding CoM motions. It should be noted that the extracted dimensions do not necessarily correspond to the true semi-axes of the nanocrystals. Owing to optical torque induced alignment, the crystal $Z$ axis tends to align with the polarization direction of the trapping laser, and the measured dimensions therefore correspond to effective particle sizes projected onto the laboratory coordinate system. Nevertheless, the differences among the three particles clearly indicate that they possess distinct and randomly distributed shapes, providing an appropriate basis for evaluating the influence of particle shape on the observed alignment behavior.


\begin{acknowledgments}
	This research is supported by the Quantum Science and Technology-National Science and Technology Major Project (Grant No. 2021ZD0302003) and the National Natural Science Foundation of China (Grant Nos. 61975101,U23A6004, 92476001, 12488301, 12034011).
\end{acknowledgments}

\bibliography{references}

@article{GonzalezLevitodynamics2021,
	title = {Levitodynamics: Levitation and control of microscopic objects in vacuum},
	volume = {374},
	url = {https://www.science.org/doi/10.1126/science.abg3027},
	doi = {10.1126/science.abg3027},
	number = {6564},
	journal = {Science},
	author = {Gonzalez-Ballestero, C. and Aspelmeyer, M. and Novotny, L. and Quidant, R. and Romero-Isart, O.},
	year = {2021},
	pages = {eabg3027}
}

@misc{JinTowards2024,
	title = {Towards real-world applications of levitated optomechanics},
	url = {https://arxiv.org/abs/2407.12496},
	doi = {10.48550/arXiv.2407.12496},
	archivePrefix = {arXiv},
	author = {Jin, Yuanbin and Shen, Kunhong and Ju, Peng and Li, Tongcang},
	year = {2024},
	eprint = {2407.12496}
}

@article{VolpeRoadmap2023,
	title = {Roadmap for optical tweezers},
	volume = {5},
	url = {https://dx.doi.org/10.1088/2515-7647/acb57b},
	doi = {10.1088/2515-7647/acb57b},
	number = {2},
	journal = {J. Phys. Photonics},
	year = {2023},
	author = {Volpe, Giovanni and Maragò, Onofrio M. and Rubinsztein-Dunlop, Halina and Pesce, Giuseppe and Stilgoe, Alexander B. and Volpe, Giorgio and Tkachenko, Georgiy and Truong, Viet Giang and Chormaic, Síle Nic and Kalantarifard, Fatemeh and Elahi, Parviz and Käll, Mikael and Callegari, Agnese and Marqués, Manuel I. and Neves, Antonio A. R. and Moreira, Wendel L. and Fontes, Adriana and Cesar, Carlos L. and Saija, Rosalba and Saidi, Abir and Beck, Paul and Eismann, Jörg S. and Banzer, Peter and Fernandes, Thales F. D. and Pedaci, Francesco and Bowen, Warwick P. and Vaippully, Rahul and Lokesh, Muruga and Roy, Basudev and Thalhammer-Thurner, Gregor and Ritsch-Marte, Monika and García, Laura Pérez and Arzola, Alejandro V. and Castillo, Isaac Pérez and Argun, Aykut and Muenker, Till M. and Vos, Bart E. and Betz, Timo and Cristiani, Ilaria and Minzioni, Paolo and Reece, Peter J. and Wang, Fan and McGloin, David and Ndukaife, Justus C. and Quidant, Romain and Roberts, Reece P. and Laplane, Cyril and Volz, Thomas and Gordon, Reuven and Hanstorp, Dag and Marmolejo, Javier Tello and Bruce, Graham D. and Dholakia, Kishan and Li, Tongcang and Brzobohatý, Oto and Simpson, Stephen H. and Zemánek, Pavel and Ritort, Felix and Roichman, Yael and Bobkova, Valeriia and Wittkowski, Raphael and Denz, Cornelia and Kumar, G. V. Pavan and Foti, Antonino and Donato, Maria Grazia and Gucciardi, Pietro G. and Gardini, Lucia and Bianchi, Giulio and Kashchuk, Anatolii V. and Capitanio, Marco and Paterson, Lynn and Jones, Philip H. and Berg-Sørensen, Kirstine and Barooji, Younes F. and Oddershede, Lene B. and Pouladian, Pegah and Preece, Daryl and Adiels, Caroline Beck and Luca, Anna Chiara De and Magazzù, Alessandro and Ciriza, David Bronte and Iatì, Maria Antonia and Swartzlander, Grover A.},
	pages = {022501}
}

@article{LiMeasurement2010,
	title = {Measurement of the Instantaneous Velocity of a Brownian Particle},
	volume = {328},
	url = {https://www.science.org/doi/10.1126/science.1189403},
	doi = {10.1126/science.1189403},
	number = {5986},
	journal = {Science},
	author = {Li, Tongcang and Kheifets, Simon and Medellin, David and Raizen, Mark G.},
	year = {2010},
	pages = {1673--1675}
}

@article{AhnUltrasensitive2020,
	title = {Ultrasensitive torque detection with an optically levitated nanorotor},
	volume = {15},
	url = {https://www.nature.com/articles/s41565-019-0605-9},
	doi = {10.1038/s41565-019-0605-9},
	number = {2},
	journal = {Nat. Nanotechnol.},
	author = {Ahn, Jonghoon and Xu, Zhujing and Bang, Jaehoon and Ju, Peng and Gao, Xingyu and Li, Tongcang},
	year = {2020},
	pages = {89--93}
}

@article{MonteiroForce2020,
	title = {Force and acceleration sensing with optically levitated nanogram masses at microkelvin temperatures},
	volume = {101},
	url = {https://link.aps.org/doi/10.1103/PhysRevA.101.053835},
	doi = {10.1103/PhysRevA.101.053835},
	number = {5},
	journal = {Phys. Rev. A},
	author = {Monteiro, Fernando and Li, Wenqiang and Afek, Gadi and Li, Chang-ling and Mossman, Michael and Moore, David C.},
	year = {2020},
	pages = {053835}
}

@article{DelicCooling2020,
	title = {Cooling of a levitated nanoparticle to the motional quantum ground state},
	volume = {367},
	url = {https://www.science.org/doi/10.1126/science.aba3993},
	doi = {10.1126/science.aba3993},
	number = {6480},
	journal = {Science},
	author = {Delić, Uroš and Reisenbauer, Manuel and Dare, Kahan and Grass, David and Vuletić, Vladan and Kiesel, Nikolai and Aspelmeyer, Markus},
	year = {2020},
	pages = {892--895}
}

@article{MagriniReal2021,
	title = {Real-time optimal quantum control of mechanical motion at room temperature},
	volume = {595},
	url = {https://www.nature.com/articles/s41586-021-03602-3},
	doi = {10.1038/s41586-021-03602-3},
	number = {7867},
	journal = {Nature},
	author = {Magrini, Lorenzo and Rosenzweig, Philipp and Bach, Constanze and Deutschmann-Olek, Andreas and Hofer, Sebastian G. and Hong, Sungkun and Kiesel, Nikolai and Kugi, Andreas and Aspelmeyer, Markus},
	year = {2021},
	pages = {373--377}
}

@article{TebbenjohannsQuantum2021,
	title = {Quantum control of a nanoparticle optically levitated in cryogenic free space},
	volume = {595},
	url = {https://www.nature.com/articles/s41586-021-03617-w},
	doi = {10.1038/s41586-021-03617-w},
	number = {7867},
	journal = {Nature},
	author = {Tebbenjohanns, Felix and Mattana, M. Luisa and Rossi, Massimiliano and Frimmer, Martin and Novotny, Lukas},
	year = {2021},
	pages = {378--382}
}

@article{JinGHz2021,
	title = {6 {GHz} hyperfast rotation of an optically levitated nanoparticle in vacuum},
	volume = {9},
	url = {https://opg.optica.org/abstract.cfm?URI=prj-9-7-1344},
	doi = {10.1364/PRJ.422975},
	number = {7},
	journal = {Photon. Res.},
	author = {Jin, Yuanbin and Yan, Jiangwei and Rahman, Shah Jee and Li, Jie and Yu, Xudong and Zhang, Jing},
	year = {2021},
	pages = {1344}
}

@article{AhnOptically2018,
	title = {Optically Levitated Nanodumbbell Torsion Balance and {GHz} Nanomechanical Rotor},
	volume = {121},
	url = {https://link.aps.org/doi/10.1103/PhysRevLett.121.033603},
	doi = {10.1103/PhysRevLett.121.033603},
	number = {3},
	journal = {Phys. Rev. Lett.},
	author = {Ahn, Jonghoon and Xu, Zhujing and Bang, Jaehoon and Deng, Yu-Hao and Hoang, Thai M. and Han, Qinkai and Ma, Ren-Min and Li, Tongcang},
	year = {2018},
	pages = {033603}
}

@article{GeraciShort2010,
	title = {Short-Range Force Detection Using Optically Cooled Levitated Microspheres},
	volume = {105},
	url = {https://link.aps.org/doi/10.1103/PhysRevLett.105.101101},
	doi = {10.1103/PhysRevLett.105.101101},
	number = {10},
	journal = {Phys. Rev. Lett.},
	author = {Geraci, Andrew A. and Papp, Scott B. and Kitching, John},
	year = {2010},
	pages = {101101}
}

@article{ChenConstraining2022,
	title = {Constraining the axion-nucleon coupling and non-{Newtonian} gravity with a levitated optomechanical device},
	volume = {106},
	url = {https://link.aps.org/doi/10.1103/PhysRevD.106.095007},
	doi = {10.1103/PhysRevD.106.095007},
	number = {9},
	journal = {Phys. Rev. D},
	author = {Chen, Lei and Liu, Jian and Zhu, Ka-Di},
	year = {2022},
	pages = {095007}
}

@article{JinQuantum2024,
	title = {Quantum control and {Berry} phase of electron spins in rotating levitated diamonds in high vacuum},
	volume = {15},
	url = {https://www.nature.com/articles/s41467-024-49175-3},
	doi = {10.1038/s41467-024-49175-3},
	number = {1},
	journal = {Nat. Commun.},
	author = {Jin, Yuanbin and Shen, Kunhong and Ju, Peng and Gao, Xingyu and Zu, Chong and Grine, Alejandro J. and Li, Tongcang},
	year = {2024},
	pages = {5063}
}

@article{NeukirchMulti2015,
	title = {Multi-dimensional single-spin nano-optomechanics with a levitated nanodiamond},
	volume = {9},
	url = {https://www.nature.com/articles/nphoton.2015.162},
	doi = {10.1038/nphoton.2015.162},
	number = {10},
	journal = {Nat. Photon.},
	author = {Neukirch, Levi P. and von Haartman, Eva and Rosenholm, Jessica M. and Nick Vamivakas, A.},
	year = {2015},
	pages = {653--657}
}

@article{MaclaurinMeasurable2012,
	title = {Measurable Quantum Geometric Phase from a Rotating Single Spin},
	volume = {108},
	url = {https://link.aps.org/doi/10.1103/PhysRevLett.108.240403},
	doi = {10.1103/PhysRevLett.108.240403},
	number = {24},
	journal = {Phys. Rev. Lett.},
	author = {Maclaurin, D. and Doherty, M. W. and Hollenberg, L. C. L. and Martin, A. M.},
	year = {2012},
	pages = {240403}
}

@article{LePhotostable2008,
	title = {Photostable Second Harmonic Generation from a Single {KTiOPO$_4$} Nanocrystal for Nonlinear Microscopy},
	volume = {4},
	url = {https://onlinelibrary.wiley.com/doi/abs/10.1002/smll.200701093},
	doi = {10.1002/smll.200701093},
	number = {9},
	journal = {Small},
	author = {Le Xuan, Loc and Zhou, Chunyuan and Slablab, Abdallah and Chauvat, Dominique and Tard, Cédric and Perruchas, Sandrine and Gacoin, Thierry and Villeval, Philippe and Roch, Jean-François},
	year = {2008},
	pages = {1332--1336}
}

@article{ZhangThree2011,
	title = {Three-Dimensional Nanostructures as Highly Efficient Generators of Second Harmonic Light},
	volume = {11},
	url = {https://pubs.acs.org/doi/10.1021/nl2033602},
	doi = {10.1021/nl2033602},
	number = {12},
	journal = {Nano Lett.},
	author = {Zhang, Yu and Grady, Nathaniel K. and Ayala-Orozco, Ciceron and Halas, Naomi J.},
	year = {2011},
	pages = {5519--5523}
}

@article{BonacinaHarmonic2020,
	title = {Harmonic generation at the nanoscale},
	volume = {127},
	url = {https://pubs.aip.org/aip/jap/article/127/23/230901/157227/Harmonic-generation-at-the-nanoscale},
	doi = {10.1063/5.0006093},
	number = {23},
	journal = {J. Appl. Phys.},
	author = {Bonacina, Luigi and Brevet, Pierre-François and Finazzi, Marco and Celebrano, Michele},
	year = {2020},
	pages = {230901}
}

@article{SandeauDefocused2007,
	title = {Defocused imaging of second harmonic generation from a single nanocrystal},
	volume = {15},
	url = {https://opg.optica.org/oe/abstract.cfm?uri=oe-15-24-16051},
	doi = {10.1364/OE.15.016051},
	number = {24},
	journal = {Opt. Express},
	author = {Sandeau, N. and Xuan, L. Le and Chauvat, D. and Zhou, C. and Roch, J.-F. and Brasselet, S.},
	year = {2007},
	pages = {16051--16060}
}

@article{YangThe2020,
	title = {The study of structure evolvement of {KTiOPO$_4$} family and their nonlinear optical properties},
	volume = {423},
	url = {https://www.sciencedirect.com/science/article/pii/S0010854520305567},
	doi = {10.1016/j.ccr.2020.213491},
	journal = {Coord. Chem. Rev.},
	author = {Yang, Fei and Wang, Lei and Huang, Ling and Zou, Guohong},
	year = {2020},
	pages = {213491}
}

@article{ChenHigh2021,
	title = {High-Performance Second Harmonic Generation ({SHG}) Materials: New Developments and New Strategies},
	volume = {54},
	url = {https://doi.org/10.1021/acs.accounts.1c00188},
	doi = {10.1021/acs.accounts.1c00188},
	number = {12},
	journal = {Acc. Chem. Res.},
	author = {Chen, Jin and Hu, Chun-Li and Kong, Fang and Mao, Jiang-Gao},
	year = {2021},
	pages = {2775--2783}
}

@article{ChenOptical2025,
	title = {Optical nonlinearities in excess of 500 through sublattice reconstruction},
	volume = {643},
	url = {https://www.nature.com/articles/s41586-025-09164-y},
	doi = {10.1038/s41586-025-09164-y},
	number = {8072},
	journal = {Nature},
	author = {Chen, Jiaye and Liu, Chang and Xi, Shibo and Tan, Shengdong and He, Qian and Liang, Liangliang and Liu, Xiaogang},
	year = {2025},
	pages = {669--674}
}

@article{TzarouchisLight2018,
	title = {Light Scattering by a Dielectric Sphere: Perspectives on the Mie Resonances},
	volume = {8},
	url = {https://www.mdpi.com/2076-3417/8/2/184},
	doi = {10.3390/app8020184},
	number = {2},
	journal = {Appl. Sci.},
	author = {Tzarouchis, Dimitrios and Sihvola, Ari},
	year = {2018},
	pages = {184}
}

@article{LiangYoctonewton2023,
	title = {Yoctonewton force detection based on optically levitated oscillator},
	volume = {3},
	url = {https://www.sciencedirect.com/science/article/pii/S2667325822003879},
	doi = {10.1016/j.fmre.2022.09.021},
	number = {1},
	journal = {Fundam. Res.},
	author = {Liang, Tao and Zhu, Shaochong and He, Peitong and Chen, Zhiming and Wang, Yingying and Li, Cuihong and Fu, Zhenhai and Gao, Xiaowen and Chen, Xinfan and Li, Nan and Zhu, Qi and Hu, Huizhu},
	year = {2023},
	pages = {57--62}
}

@article{HoangElectron2016,
	title = {Electron spin control of optically levitated nanodiamonds in vacuum},
	volume = {7},
	url = {https://www.nature.com/articles/ncomms12250},
	doi = {10.1038/ncomms12250},
	number = {1},
	journal = {Nat. Commun.},
	author = {Hoang, Thai M. and Ahn, Jonghoon and Bang, Jaehoon and Li, Tongcang},
	year = {2016},
	pages = {12250}
}

@article{Jinoptically2018,
	title = {Optically levitated nanosphere with high trapping frequency},
	volume = {61},
	url = {https://doi.org/10.1007/s11433-018-9230-6},
	doi = {10.1007/s11433-018-9230-6},
	number = {11},
	journal = {Sci. China-Phys. Mech. Astron.},
	author = {Jin, YuanBin and Yu, XuDong and Zhang, Jing},
	year = {2018},
	pages = {114221}
}

@article{BangFive2020,
	title = {Five-dimensional cooling and nonlinear dynamics of an optically levitated nanodumbbell},
	volume = {2},
	url = {https://link.aps.org/doi/10.1103/PhysRevResearch.2.043054},
	doi = {10.1103/PhysRevResearch.2.043054},
	number = {4},
	journal = {Phys. Rev. Res.},
	author = {Bang, Jaehoon and Seberson, T. and Ju, Peng and Ahn, Jonghoon and Xu, Zhujing and Gao, Xingyu and Robicheaux, F. and Li, Tongcang},
	year = {2020},
	pages = {043054}
}

@article{SomersMeasurement2018,
	title = {Measurement of the {Casimir} torque},
	volume = {564},
	url = {https://www.nature.com/articles/s41586-018-0777-8},
	doi = {10.1038/s41586-018-0777-8},
	number = {7736},
	journal = {Nature},
	author = {Somers, David A. T. and Garrett, Joseph L. and Palm, Kevin J. and Munday, Jeremy N.},
	year = {2018},
	pages = {386--389}
}

@article{XuEnhancement2021,
	title = {Enhancement of rotational vacuum friction by surface photon tunneling},
	volume = {10},
	url = {https://www.degruyterbrill.com/document/doi/10.1515/nanoph-2020-0391/html},
	number = {1},
	journal = {Nanophotonics},
	author = {Xu, Zhujing and Jacob, Zubin and Li, Tongcang},
	year = {2021},
	pages = {537--543}
}

@article{BehelControlled2024,
	title = {Controlled Second Harmonic Generation with Optically Trapped Lithium Niobate Nanoparticles},
	volume = {24},
	url = {https://doi.org/10.1021/acs.nanolett.4c00353},
	doi = {10.1021/acs.nanolett.4c00353},
	number = {19},
	journal = {Nano Lett.},
	author = {Behel, Zacharie and Mugnier, Yannick and Le Dantec, Ronan and Chevolot, Yann and Monnier, Virginie and Brevet, Pierre-François},
	year = {2024},
	pages = {5699--5704},
}

@article{WinstoneOptical2022,
	title = {Optical Trapping of High-Aspect-Ratio {NaYF} Hexagonal Prisms for kHz-{MHz} Gravitational Wave Detectors},
	volume = {129},
	url = {https://link.aps.org/doi/10.1103/PhysRevLett.129.053604},
	doi = {10.1103/PhysRevLett.129.053604},
	number = {5},
	journal = {Phys. Rev. Lett.},
	author = {Winstone, George and Wang, Zhiyuan and Klomp, Shelby and Felsted, Robert G. and Laeuger, Andrew and Gupta, Chaman and Grass, Daniel and Aggarwal, Nancy and Sprague, Jacob and Pauzauskie, Peter J. and Larson, Shane L. and Kalogera, Vicky and Geraci, Andrew A.},
	year = {2022},
	pages = {053604}
}

@article{DelordSpin2020,
	title = {Spin-cooling of the motion of a trapped diamond},
	volume = {580},
	url = {https://www.nature.com/articles/s41586-020-2133-z},
	doi = {10.1038/s41586-020-2133-z},
	number = {7801},
	journal = {Nature},
	author = {Delord, T. and Huillery, P. and Nicolas, L. and Hétet, G.},
	year = {2020},
	pages = {56--59}
}

@article{RahmanLaser2017,
	title = {Laser refrigeration, alignment and rotation of levitated {Yb3}+:{YLF} nanocrystals},
	volume = {11},
	url = {https://www.nature.com/articles/s41566-017-0005-3},
	doi = {10.1038/s41566-017-0005-3},
	number = {10},
	journal = {Nat. Photon.},
	author = {Rahman, A. T. M. Anishur and Barker, P. F.},
	year = {2017},
	pages = {634--638}
}

@article{SticklerQuantum2021,
	title = {Quantum rotations of nanoparticles},
	volume = {3},
	url = {https://www.nature.com/articles/s42254-021-00335-0},
	doi = {10.1038/s42254-021-00335-0},
	number = {8},
	journal = {Nat. Rev. Phys.},
	author = {Stickler, Benjamin A. and Hornberger, Klaus and Kim, M. S.},
	year = {2021},
	pages = {589--597}
}

@article{TroyerQuantum2026,
	title = {Quantum ground-state cooling of two librational modes of a nanorotor},
	volume = {22},
	url = {https://www.nature.com/articles/s41567-026-03219-1},
	doi = {10.1038/s41567-026-03219-1},
	number = {4},
	journal = {Nat. Phys.},
	publisher = {Nature Publishing Group},
	author = {Troyer, Stephan and Fechtel, Florian and Hummer, Lorenz and Rudolph, Henning and Stickler, Benjamin A. and Delić, Uroš and Arndt, Markus},
	year = {2026},
	pages = {584--590}
}

@article{PontinSimultaneous2023,
	title = {Simultaneous cavity cooling of all six degrees of freedom of a levitated nanoparticle},
	volume = {19},
	url = {https://www.nature.com/articles/s41567-023-02006-6},
	doi = {10.1038/s41567-023-02006-6},
	number = {7},
	journal = {Nat. Phys.},
	author = {Pontin, A. and Fu, H. and Toroš, M. and Monteiro, T. S. and Barker, P. F.},
	year = {2023},
	pages = {1003--1008},
}

@article{FishmanPhoton2023,
	title = {Photon-Emission-Correlation Spectroscopy as an Analytical Tool for Solid-State Quantum Defects},
	volume = {4},
	url = {https://link.aps.org/doi/10.1103/PRXQuantum.4.010202},
	doi = {10.1103/PRXQuantum.4.010202},
	number = {1},
	journal = {PRX Quantum},
	author = {Fishman, Rebecca E.K. and Patel, Raj N. and Hopper, David A. and Huang, Tzu-Yung and Bassett, Lee C.},
	year = {2023},
	pages = {010202}
}

@article{YuExperimental2014,
	title = {Experimental measurement of covariance matrix of two-mode entangled state},
	volume = {57},
	url = {https://doi.org/10.1007/s11433-014-5448-7},
	doi = {10.1007/s11433-014-5448-7},
	number = {5},
	journal = {Sci. China Phys. Mech. Astron.},
	author = {Yu, XuDong and Li, Wei and Jin, YuanBin and Zhang, Jing},
	year = {2014},
	pages = {875--879}
}

@article{WuGeneration1986,
	title = {Generation of Squeezed States by Parametric Down Conversion},
	volume = {57},
	url = {https://link.aps.org/doi/10.1103/PhysRevLett.57.2520},
	doi = {10.1103/PhysRevLett.57.2520},
	number = {20},
	journal = {Phys. Rev. Lett.},
	author = {Wu, Ling-An and Kimble, H. J. and Hall, J. L. and Wu, Huifa},
	year = {1986},
	pages = {2520--2523}
}

@article{TrojekOptical2012,
	author = {Jan Trojek and Luk\'{a}\v{s} Chv\'{a}tal and Pavel Zem\'{a}nek},
	journal = {J. Opt. Soc. Am. A},
	number = {7},
	pages = {1224--1236},
	publisher = {Optica Publishing Group},
	title = {Optical alignment and confinement of an ellipsoidal nanorod in optical tweezers: a theoretical study},
	volume = {29},
	month = {Jul},
	year = {2012},
	url = {https://opg.optica.org/josaa/abstract.cfm?URI=josaa-29-7-1224},
	doi = {10.1364/JOSAA.29.001224}
}

@article{JinImaging2021,
	title = {Imaging the dipole scattering of an optically levitated dielectric nanoparticle},
	volume = {119},
	url = {https://doi.org/10.1063/5.0053008},
	doi = {10.1063/5.0053008},
	number = {2},
	journal = {Appl. Phys. Lett.},
	author = {Jin, Yuanbin and Yan, Jiangwei and Rahman, Shah Jee and Yu, Xudong and Zhang, Jing},
	year = {2021},
	pages = {021106},
}


%
%
%
%

\newpage
\onecolumngrid
\appendix
\renewcommand{\figurename}{Supplemental Fig.}
\renewcommand{\tablename}{Supplemental Table}
\setcounter{figure}{0}
\setcounter{table}{0}

\section*{Supplementary Information}

\section{Experimental setup}

\begin{figure}
	\centering\includegraphics[width=14cm]{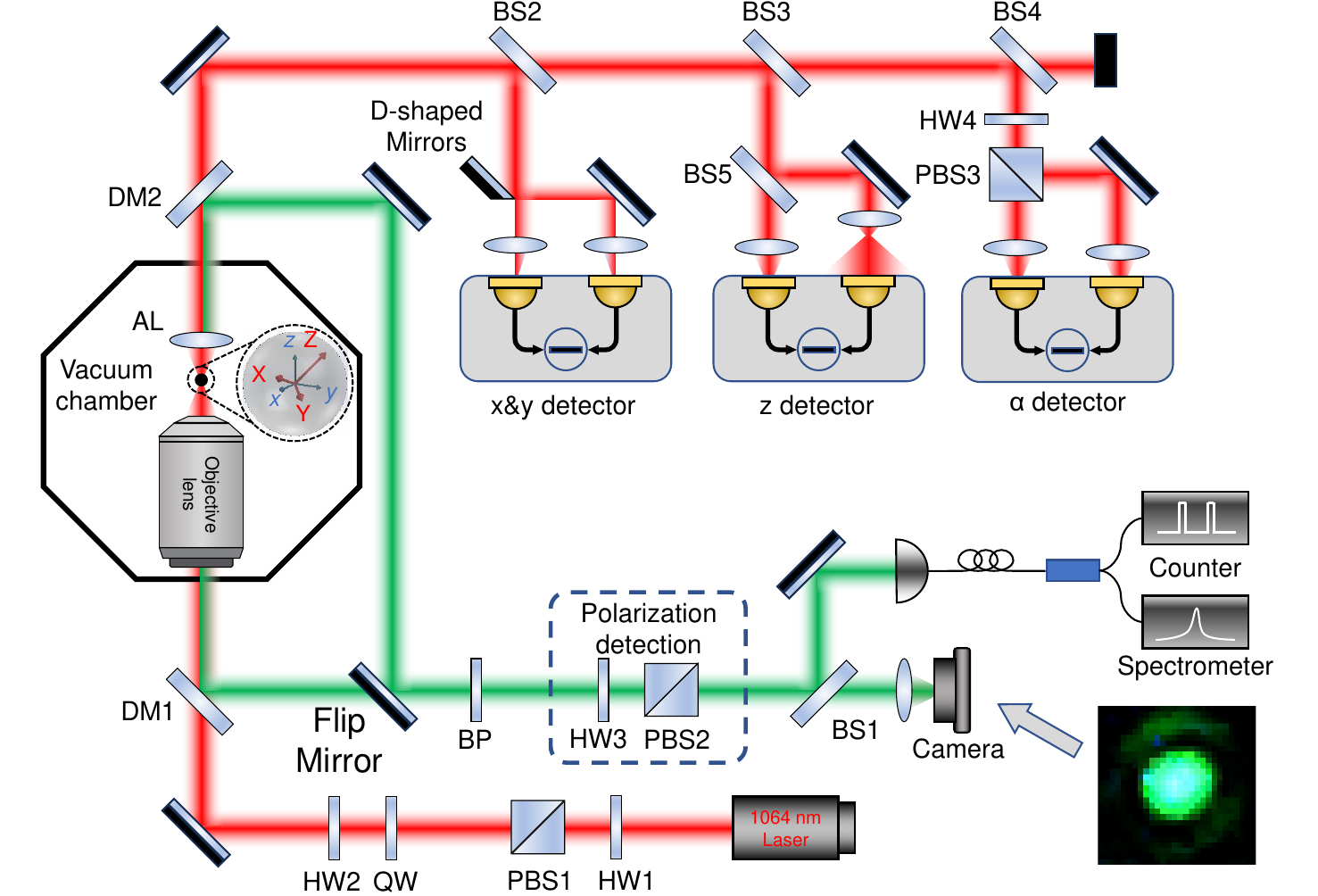}
	\caption{\label{fig:s1} Experimental setup for optical levitation and SHG characterization of a KTP nanocrystal. A 1064 nm laser with independently controllable power and polarization is tightly focused by a high numerical aperture objective lens (NA = 0.95) to optically levitate a KTP nanocrystal in vacuum. The forward-scattered trapping light is collected by an aspheric lens (AL) and directed to a motion detection system consisting of three channels for monitoring the nanoparticle center-of-mass (CoM) motion along the $x$-$y$ and $z$ directions, as well as its librational motion ($\alpha$). The backward-propagating second-harmonic generation (SHG) signal is collected by the objective lens and characterized using a color camera, a single-photon counter, and a spectrometer. Polarization-resolved measurements of the SHG emission are performed using a half-wave plate (HW3) and a polarizing beam splitter (PBS2). In addition, the forward-propagating SHG signal collected by the aspheric lens is redirected into the same detection path, enabling comparison of the forward and backward SHG emission and characterization of the three-dimensional radiation pattern. The inset shows a camera image of the SHG emission from a levitated KTP nanocrystal. DM1-2, dichroic mirrors; HW1-4, half-wave plates; QW, quarter-wave plate; BP, bandpass filter; BS1-5, beam splitters. Two coordinate systems are defined throughout this work: the laboratory frame ($xyz$) and the crystallographic coordinate system ($XYZ$).}
\end{figure}

Supplemental Fig. \ref{fig:s1} shows the complete experimental setup. A 1064 nm laser beam from a solid-state laser is first passed through a half-wave plate (HW1) and a polarizing beam splitter (PBS1) for power control. The polarization state of the beam is subsequently adjusted using a quarter-wave plate (QW) and a second half-wave plate (HW2). The beam is then tightly focused by a high numerical aperture objective lens (NA = 0.95) to form an optical trap for a KTP nanocrystal levitation in vacuum. The trapping laser simultaneously serves as the fundamental field for second-harmonic generation (SHG).

The forward-scattered 1064 nm laser is collected by an aspheric lens (AL) and directed to a balanced homodyne detection system for monitoring the center-of-mass (CoM) and librational motions of the levitated nanocrystal. The detection system consists of three independent channels corresponding to the $x$-$y$ CoM motion, the $z$ CoM motion, and the librational motion ($\alpha$), respectively. A detailed description of the detection scheme can be found in Ref. \cite{JinTowards2024}.

When the tightly focused fundamental field interacts with the levitated KTP nanocrystal, SHG is generated and radiated into three-dimensional space. The backward-propagating SHG signal is collected by the objective lens and separated from the fundamental laser using a dichroic mirror (DM1). In addition, the forward-propagating SHG signal is collected by the aspheric lens and redirected into the same detection path as the backward-collected SHG signal. In both cases, a bandpass filter (BP, center wavelength 532 nm, FWHM bandwidth 10 nm) is used to suppress the residual 1064 nm trapping light. The filtered SHG signal is directed to a color camera for real-time imaging, a single-photon counter for intensity measurements, and a spectrometer for spectral characterization.

Because the dimensions of the levitated KTP nanocrystals are substantially smaller than the fundamental laser wavelength, the phase-matching condition required for bulk nonlinear crystals is no longer relevant. In this subwavelength regime, the SHG emission can be approximated as radiation from a dipole source. To verify this picture experimentally, we compare the SHG signals collected in the forward and backward directions. Figure 2(a) of the main text shows the measured SHG photon counting rates for linearly polarized fundamental laser. The objective lens and aspheric lens have numerical apertures of 0.95 and 0.50, respectively. 

\added{We estimated the SHG conversion efficiency from the backward-collected photon-counting signal. Assuming that the SHG emission follows the dipole radiation pattern, with the dipole axis perpendicular to the optical axis, the spatial distribution of the light scattering in free space can be written as \cite{JinImaging2021}
	\begin{equation}
		I = \frac{3}{{8\pi {r^2}}}{I_{tot}}{\sin ^2}\Theta  = \frac{3}{{8\pi {r^2}}}{I_{tot}}\left[ {1 - {{\left( {\sin {\theta _k}\cos {\varphi _k}} \right)}^2}} \right],
	\end{equation}
	where $I_tot$ is total radiation SHG intensity, $\theta_k$ is the polar angle with respect to the $z$ axis, and $\varphi_k$ is the azimuthal angle measured from the $x$ axis in the $xy$-plane. Therefore, the collection efficiency of an objective with a maximum collection angle $\theta_m$ is
	\begin{equation}
		\begin{array}{l}
			\eta  = \frac{{{I_{col}}}}{{{I_{tot}}}} = \int_0^{{\theta _m}} {\int_0^{2\pi } {\frac{3}{{8\pi {r^2}}}\left[ {1 - {{\left( {\sin {\theta _k}\cos {\varphi _k}} \right)}^2}} \right]{r^2}\sin {\theta _k}d{\theta _k}d{\varphi _k}} } \\
			= \frac{1}{8}\left( {4 - 3\cos {\theta _m} - {{\cos }^3}{\theta _m}} \right)
		\end{array},
	\end{equation}
	with ${\theta _m} = \arcsin \left( {NA} \right)$. For the NA = 0.95 objective lens used for backward collection, this gives a collection efficiency of approximately 37.9\%. Taking into account the objective transmission at 532 nm (0.64), the fiber-coupling efficiency (0.41), and the quantum efficiency of the single-photon detector (0.51), the overall detection efficiency is approximately 5.1\%. At a fundamental power of 120 mW, the measured backward photon counting rate is about $2.28 \times 10^5$/s. The corresponding total SHG photon generation rate is estimated as $4.47 \times 10^6$/s. Since each SHG photon corresponds to the conversion of two fundamental photons, the corresponding SHG conversion efficiency is approximately $1.4 \times 10^{-11}$.}

\added{For the forward collected SHG signal, an aspheric lens with $NA = 0.5$ is used. In addition to its smaller collection solid angle, the forward channel suffers from substantially poorer spatial-mode matching and additional optical losses. The measured forward photon counting rate of $7.94 \times 10^2/s$ corresponds to an effective overall forward detection efficiency of approximately 0.18\%.}

\deleted{The overall collection efficiency of the backward detection path is approximately twice that of the forward detection path, primarily due to differences in optical transmission. After correcting for these collection efficiencies, the inferred SHG powers emitted into the forward and backward directions are nearly identical. This observation is consistent with the dipole radiation pattern expected for SHG from a subwavelength nonlinear nanocrystal.}

\section{Second-order nonlinear susceptibility tensor of KTP crystal}

The KTP nanocrystal is non-centrosymmetric orthorhombic crystals with characteristic dimensions of approximately 100 nm used in this work. The SHG process is governed by the second-order nonlinear polarization induced by the fundamental field. For an incident electric field ${\bf{E}} = \left( {{E_X},{E_Y},{E_Z}} \right)$ with a frequency of $\omega$, the induced second-order polarization can be written as
\begin{equation}
	P_i^{\left( 2 \right)} = {\varepsilon _0}\sum\limits_{j,k} {\chi _{ijk}^{\left( 2 \right)}{E_j}{E_k}},
\end{equation}
where $\varepsilon_0$ is the vacuum permittivity and $\chi_{ijk}^{(2)}$ is the second-order susceptibility tensor. The indices $i$, $j$, and $k$ refer to the principle axes ($X$, $Y$, and $Z$) of the KTP crystal.

Following the standard contracted notation for second-order nonlinear optics, the susceptibility tensor is expressed in terms of the reduced nonlinear coefficient matrix $d_{il}$ according to ${d_{il}} = \chi _{ijk}^{\left( 2 \right)}/2$, where the pair of indices $(j,k)$ is contracted into the single index $l$ following the convention summarized in Supplemental Table \ref{tab:s1}.

\begin{table}
	\caption{Contracted notation convention relating the tensor indices $\left( j,k \right)$ of the second-order susceptibility tensor $\chi_{ijk}^{\left( 2 \right)}$ to the single index $l$ in the reduced nonlinear coefficient matrix ${d_{il}}$.}
	\centering
	\label{tab:s1} 
	\begin{tabular}{ccccccc}
		\hline
		index $l$ & 1 & 2 & 3 & 4 & 5 & 6 \\
		\hline
		index $\left( j,k \right)$ & $\left( x,x \right)$ & $\left( y,y \right)$ & $\left( z,z \right)$ & $\left( y,z \right)$ & $\left( x,z \right)$ & $\left( x,y \right)$ \\
		\hline
	\end{tabular}
\end{table}

For KTP crystals belonging to the orthorhombic point group $mm2$, the reduced nonlinear coefficient matrix is
\begin{equation}
	{d_{il}} = \left( {\begin{array}{*{20}{c}}
			0&0&0&0&{{d_{15}}}&0\\
			0&0&0&{{d_{24}}}&0&0\\
			{{d_{31}}}&{{d_{32}}}&{{d_{33}}}&0&0&0
	\end{array}} \right).
\end{equation}
The nonlinear coefficients provided by the manufacturer are ${d_{31}} = 6.5$ pm/V, ${d_{32}} = 5$ pm/V, ${d_{33}} = 13.7$ pm/V, ${d_{24}} = 7.6$ pm/V, and ${d_{15}} = 6.1$ pm/V. Among these coefficients, $d_{33}$ is the largest and dominates the nonlinear response when the fundamental electric field contains a component along the crystal $Z$ axis. This strong anisotropy is responsible for the pronounced polarization dependence of the SHG emission observed in the main text and plays a key role in the optical alignment behavior of the levitated KTP nanocrystals.

\section{Calculation of SHG emission}
To analyze the polarization dependence of the SHG, we assume that the major axis of the polarization ellipse of the fundamental laser is aligned with the crystal $Z$ axis. Under this condition, the propagation direction of the laser lies in the $XY$ plane and forms an angle $\varphi$ with respect to the crystal $X$ axis. The electric field of the fundamental laser can be expressed in the crystal coordinate system as
\begin{equation}
	\begin{array}{l}
		{E_X} = \pm i{E_0}\sin \theta \cos \varphi {e^{ - i\omega t}} \\
		{E_Y} = \pm i{E_0}\sin \theta \sin \varphi {e^{ - i\omega t}} \\
		{E_Z} = {E_0}\cos \theta {e^{ - i\omega t}}
	\end{array},
\end{equation}
where the nanocrystal is assumed to be trapped at the focal point of the trapping laser and $E_0$ is the amplitude of the fundamental electric field. The parameter $\theta$ characterizes the polarization ellipticity of the fundamental field, with $\cos \theta  = {E_{0Z}}/{E_0}$, where $E_{0Z}$ is the projection of the electric field onto the crystal $Z$ axis. The positive and negative signs correspond to left- and right-handed polarization, respectively.

Substituting the electric field components into the second-order nonlinear polarization tensor obtains the SHG polarization components along the three directions,
\begin{equation}
	\begin{array}{l}
		{P_Z} = {\varepsilon _0}\chi _{ZZZ}^{\left( 2 \right)}E_Z^2 + {\varepsilon _0}\chi _{ZXX}^{\left( 2 \right)}E_X^2 + {\varepsilon _0}\chi _{ZYY}^{\left( 2 \right)}E_Y^2 \\
		= 2{\varepsilon _0}{d_{33}}{\left( {{E_0}\cos \theta {e^{ - i\omega t}}} \right)^2} + 2{\varepsilon _0}{d_{31}}{\left( { \pm i{E_0}\sin \theta \cos \varphi {e^{ - i\omega t}}} \right)^2} + 2{\varepsilon _0}{d_{32}}{\left( { \pm i{E_0}\sin \theta \sin \varphi {e^{ - i\omega t}}} \right)^2}\\
		= 2{\varepsilon _0}E_0^2{e^{ - 2i\omega t}}\left[ {{d_{33}}{{\cos }^2}\theta  - \left( {{d_{31}}{{\cos }^2}\varphi  + {d_{32}}{{\sin }^2}\varphi } \right){{\sin }^2}\theta } \right]
	\end{array},
\end{equation}
\begin{equation}
	\begin{array}{l}
		{P_X} = {\varepsilon _0}\chi _{XXZ}^{\left( 2 \right)}{E_X}{E_Z} + {\varepsilon _0}\chi _{XZX}^{\left( 2 \right)}{E_Z}{E_X}\\
		= 4{\varepsilon _0}{d_{15}}\left( { \pm i{E_0}\sin \theta \cos \varphi {e^{ - i\omega t}}} \right)\left( {{E_0}\cos \theta {e^{ - i\omega t}}} \right)\\
		= 2{\varepsilon _0}E_0^2{e^{ - 2i\omega t}}\left( { \pm 2i{d_{15}}\sin \theta \cos \theta \cos \varphi } \right)
	\end{array},
\end{equation}
\begin{equation}
	\begin{array}{l}
		{P_Y} = {\varepsilon _0}\chi _{YYZ}^{\left( 2 \right)}{E_Y}{E_Z} + {\varepsilon _0}\chi _{YZY}^{\left( 2 \right)}{E_Z}{E_Y}\\
		= 4{\varepsilon _0}{d_{24}}\left( { \pm i{E_0}\sin \theta \sin \varphi {e^{ - i\omega t}}} \right)\left( {{E_0}\cos \theta {e^{ - i\omega t}}} \right)\\
		= 2{\varepsilon _0}E_0^2{e^{ - 2i\omega t}}\left( { \pm 2i{d_{24}}\sin \theta \cos \theta \sin \varphi } \right)
	\end{array}.
\end{equation}
The total nonlinear polarization is then obtained by summing the contributions along the $X$, $Y$, and $Z$ axes,
\begin{equation}
	\begin{array}{l}
		{P^{\left( 2 \right)}} = {P_X}{{\bf{e}}_X} + {P_Y}{{\bf{e}}_Y} + {P_Z}{{\bf{e}}_Z}\\
		= 2{\varepsilon _0}E_0^2{e^{ - 2i\omega t}}\left[ {{d_{33}}{{\cos }^2}\theta  - \left( {{d_{31}}{{\cos }^2}\varphi  + {d_{32}}{{\sin }^2}\varphi } \right){{\sin }^2}\theta } \right]{{\bf{e}}_Z}\\
		+ 2{\varepsilon _0}E_0^2{e^{ - 2i\omega t}}\left( { \pm 2i{d_{15}}\sin \theta \cos \theta \cos \varphi } \right){{\bf{e}}_X} + 2{\varepsilon _0}E_0^2{e^{ - 2i\omega t}}\left( { \pm 2i{d_{24}}\sin \theta \cos \theta \sin \varphi } \right){{\bf{e}}_Y}\\
		= 2{\varepsilon _0}E_0^2{e^{ - 2i\omega t}}\left[ {{A_Z}{{\bf{e}}_Z} \pm i\left( {{A_X}{{\bf{e}}_X} + {A_Y}{{\bf{e}}_Y}} \right)} \right]
	\end{array},
\end{equation}
where
\begin{equation}
	\begin{array}{l}
		{A_Z} = \left[ {{d_{33}}{{\cos }^2}\theta  - \left( {{d_{31}}{{\cos }^2}\varphi  + {d_{32}}{{\sin }^2}\varphi } \right){{\sin }^2}\theta } \right]\\
		{A_X} = 2{d_{15}}\sin \theta \cos \theta \cos \varphi \\
		{A_Y} = 2{d_{24}}\sin \theta \cos \theta \sin \varphi 
	\end{array},
\end{equation}
where $\mathbf{e}_X$, $\mathbf{e}_Y$, and $\mathbf{e}_Z$ are the unit vectors along the corresponding crystal axes.

Since the levitated KTP nanocrystal is much smaller than the optical wavelength, the generated SHG can be modeled as radiation from a nonlinear dipole source. Consequently, the SHG intensity is proportional to the square of the induced second-order polarization,
\begin{equation}
	\begin{array}{l}
		I \propto {\left| {{P^{\left( 2 \right)}}} \right|^2} = 4\varepsilon _0^2E_0^4\left\{ \begin{array}{l}
			{\left[ {{d_{33}}{{\cos }^2}\theta  - \left( {{d_{31}}{{\cos }^2}\varphi  + {d_{32}}{{\sin }^2}\varphi } \right){{\sin }^2}\theta } \right]^2}\\
			+ {\left( { \pm 2i{d_{15}}\sin \theta \cos \theta \cos \varphi } \right)^2} + {\left( { \pm 2i{d_{24}}\sin \theta \cos \theta \sin \varphi } \right)^2}
		\end{array} \right\}\\
		= 4\varepsilon _0^2E_0^4\left\{ \begin{array}{l}
			{\left[ {{d_{33}}{{\cos }^2}\theta  - \left( {{d_{31}}{{\cos }^2}\varphi  + {d_{32}}{{\sin }^2}\varphi } \right){{\sin }^2}\theta } \right]^2}\\
			+ \left( {d_{15}^2{{\cos }^2}\varphi  + d_{24}^2{{\sin }^2}\varphi } \right)4{\sin ^2}\theta {\cos ^2}\theta 
		\end{array} \right\}
	\end{array}.
\end{equation}
To comparison with the experimental data, we introduce the parameter $x = \tan \theta  = \frac{b}{a}$, which corresponds to the ratio of the minor and major axes of the polarization ellipse. Using the relations
\begin{equation}
	{\cos ^2}\theta  = \frac{1}{{1 + {x^2}}}, 
	{\sin ^2}\theta  = \frac{{{x^2}}}{{1 + {x^2}}},
\end{equation}
the SHG intensity can be expressed as a function of the fundamental laser polarization ellipticity,
\begin{equation}
	\begin{array}{l}
		y = {\left[ {{d_{33}}{{\cos }^2}\theta  - \left( {{d_{31}}{{\cos }^2}\varphi  + {d_{32}}{{\sin }^2}\varphi } \right){{\sin }^2}\theta } \right]^2} + \left( {d_{15}^2{{\cos }^2}\varphi  + d_{24}^2{{\sin }^2}\varphi } \right)4{\sin ^2}\theta {\cos ^2}\theta \\
		= {\left[ {{d_{33}}\frac{1}{{1 + {x^2}}} - \left( {{d_{31}}{{\cos }^2}\varphi  + {d_{32}}{{\sin }^2}\varphi } \right)\frac{{{x^2}}}{{1 + {x^2}}}} \right]^2} + \left( {d_{15}^2{{\cos }^2}\varphi  + d_{24}^2{{\sin }^2}\varphi } \right)4\frac{{{x^2}}}{{1 + {x^2}}}\frac{1}{{1 + {x^2}}}\\
		= {\left( {\frac{1}{{1 + {x^2}}}} \right)^2}\left[ \begin{array}{l}
			d_{33}^2 + {\left( {{d_{31}}{{\cos }^2}\varphi  + {d_{32}}{{\sin }^2}\varphi } \right)^2}{x^4} - 2{d_{33}}\left( {{d_{31}}{{\cos }^2}\varphi  + {d_{32}}{{\sin }^2}\varphi } \right){x^2}\\
			+ 4\left( {d_{15}^2{{\cos }^2}\varphi  + d_{24}^2{{\sin }^2}\varphi } \right){x^2}
		\end{array} \right]\\
		= {\left( {\frac{1}{{1 + {x^2}}}} \right)^2}\left\{ \begin{array}{l}
			d_{33}^2 + {\left( {{d_{31}}{{\cos }^2}\varphi  + {d_{32}}{{\sin }^2}\varphi } \right)^2}{x^4}\\
			+ \left[ {4\left( {d_{15}^2{{\cos }^2}\varphi  + d_{24}^2{{\sin }^2}\varphi } \right) - 2{d_{33}}\left( {{d_{31}}{{\cos }^2}\varphi  + {d_{32}}{{\sin }^2}\varphi } \right)} \right]{x^2}
		\end{array} \right\}\\
		= {\left( {\frac{1}{{1 + {x^2}}}} \right)^2}\left\{ \begin{array}{l}
			d_{33}^2 + {\left( {{d_{31}}{{\cos }^2}\varphi  + {d_{32}}{{\sin }^2}\varphi } \right)^2}{x^4}\\
			+ \left[ {\left( {4d_{15}^2 - 2{d_{33}}{d_{31}}} \right){{\cos }^2}\varphi  + \left( {4d_{24}^2 - 2{d_{33}}{d_{32}}} \right){{\sin }^2}\varphi } \right]{x^2}
		\end{array} \right\}
	\end{array}.
\end{equation}

The resulting expression is used to fit the experimental data shown in Fig. 3(c) of the main text, where the polarization state of the fundamental laser is continuously varied from circular ($\tan \theta = -1 $) to linear ($\tan \theta = 0 $) and back to circular ($\tan \theta = +1 $) polarization. The agreement between theory and experiment confirms the validity of the model. The fit shows that the SHG intensity under linearly polarization is approximately five times larger than that obtained under circularly polarization. This behavior originates from the strong anisotropy of the nonlinear susceptibility tensor of KTP, in particular the dominant contribution of the $d_{33}$ coefficient when the electric field is aligned with the crystal $Z$ axis.

\section{\added{Second-order intensity-correlation measurement}}

\added{The normalized second-order intensity-correlation function is defined as
	\begin{equation}
		{g^{\left( 2 \right)}}\left( \tau  \right) = \frac{{\left\langle {I\left( t \right)I\left( {t + \tau } \right)} \right\rangle }}{{{{\left\langle {I\left( t \right)} \right\rangle }^2}}},
	\end{equation}
	where $I(t)$ is the detected SHG intensity and $\tau$ is the time delay. Experimentally, the collected SHG signal is split into two detection channels by a 50:50 fiber splitter and detected by two single photon detectors. The photon arrival times from the two channels are recorded and used to construct the coincidence histogram as a function of the relative time delay. The measured counts are then normalized to obtain $	{g^{\left( 2 \right)}}\left( \tau  \right)$. The resulting $	{g^{\left( 2 \right)}}\left( \tau  \right)$ remains close to unity, consistent with the photon statistics expected for coherently generated SHG.}

\section{Polarization analysis of SHG emission}

\begin{figure}
	\centering\includegraphics[width=14cm]{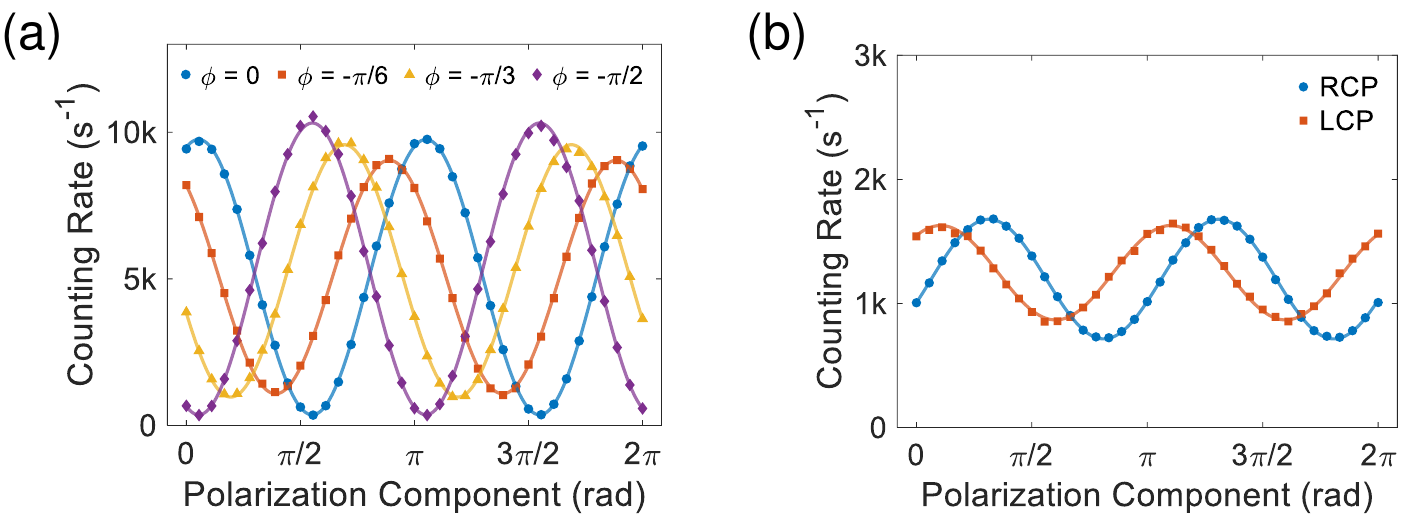}
	\caption{\label{fig:s2} Cartesian-coordinate representation of the polarization-resolved SHG measurements shown in Fig. 3(b) and (d) of the main text. (a) Polarization-resolved SHG intensity as a function of analyzer angle for fundamental laser polarization orientations of 0 (blue circles), $-\pi/6$ (red squares), $-\pi/3$ (orange triangles), and $-\pi/2$ (purple diamonds) relative to the $x$ axis. The solid curves are sinusoidal fits to the experimental data, confirming that the SHG emission is linearly polarized and aligned with the polarization of the fundamental laser. (b) Polarization-resolved SHG intensity as a function of analyzer angle under right-circularly polarized (blue circles) and left-circularly polarized (red squares) fundamental laser. The solid curves are sinusoidal fits. The measured modulation indicates an elliptically polarized SHG emission, consistent with the polarization analysis presented in the main text.}
\end{figure}

To further illustrate the polarization properties of the SHG emission, the data presented in Fig. 3(b) and (d) of the main text are replotted in Cartesian coordinates in Supplemental Fig. \ref{fig:s2}. This representation more clearly reveals the sinusoidal dependence of the detected SHG intensity on the analyzer angle.

To analyze the polarization state of the SHG emission, a half-wave plate (HW3) and a polarizing beam splitter (PBS2) are inserted into the detection path. Supplemental Fig. \ref{fig:s2}(a) shows the polarization-resolved SHG intensity as a function of the analyzer angle for four different orientations of the linearly polarized fundamental laser (blue circles: 0, red squares: $-\pi/6$, orange triangles: $-\pi/3$, and purple diamonds: $-\pi/2$). The solid curves are sinusoidal fits to the experimental data, which confirms that the SHG emission is linearly polarized and that its polarization direction follows that of the fundamental laser.

Supplemental Fig. \ref{fig:s2}(b) presents the corresponding measurements for right-circularly polarized (blue circles) and left-circularly polarized (red squares) fundamental laser. The solid curves are sinusoidal fits. In contrast to the linear polarization observed under linearly polarized fundamental laser, the SHG generated by circularly polarized fundamental laser exhibits an elliptical polarization state, consistent with the theoretical analysis presented in section 3 and the results shown in Fig. 3(d) of the main text.

\section{Optical torque and librational dynamics of levitated KTP nanocrystals}

The experimental results presented in the main text demonstrate that the levitated KTP nanocrystal continuously reorients itself to follow the polarization of the fundamental laser. This behavior is driven by polarization-dependent optical torque. The optical torque exerted by the trapping field $\mathbf{E}$ on the levitated nanocrystal can be expressed as
\begin{equation}
	{M} = \frac{1}{2}{\mathop{\rm Re}\nolimits} \left( {{\bf{p}}^* \times {\bf{E}}} \right),
\end{equation}
where $\mathbf{p}$ is the induced dipole moment. For the optical anisotropy of the KTP crystal, the torque generated by the linear optical response of the nanocrystal is substantially larger than that arising from nonlinear optical processes. Consequently, the induced dipole moment can be decomposed into two contributions associated with the shape anisotropy and the refractive index anisotropy, 
\begin{equation}
	{\bf{p}} = {{\bf{p}}_{s}} + {\bf{p}}_{r}^{\left( 1 \right)},
\end{equation}
In the following sections, we evaluate these two torque contributions separately and compare their predicted librational dynamics with the experimental observations.

\subsection{Optical torque induced by the shape anisotropy}

\begin{figure}
	\centering\includegraphics[width=14cm]{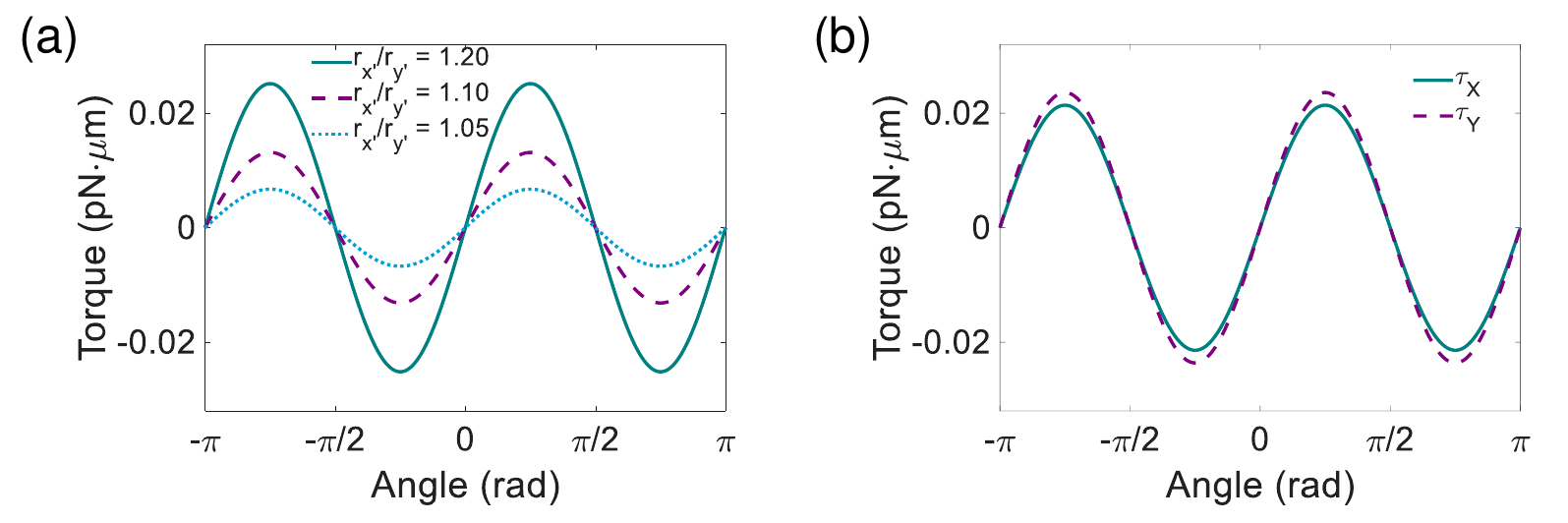}
	\caption{\label{fig:s3} Optical-torque alignment of levitated KTP nanocrystals. (a) Optical torque arising from shape anisotropy as a function of the angle between the major axis of the ellipsoidal particle and the polarization direction of the linearly polarized fundamental laser. The green solid, purple dashed, and cyan dotted curves correspond to particles with aspect ratios (long axis/short axis) of 1.20, 1.10, and 1.05, respectively, while maintaining a constant particle volume. (b) Optical torque arising from the refractive index anisotropy of KTP as a function of the angle between the crystal $Z$ axis and the polarization direction of the fundamental laser. The green solid and purple dashed curves correspond to rotations along the $X$ and $Y$ axes, respectively.}
\end{figure}

We first consider the optical torque arising solely from the shape anisotropy of an ellipsoidal particle. The induced dipole polarization can be written as
\begin{equation}
	{{\bf{p}}_s} = {\alpha _{s,x'}}{E_{x'}}{{\bf{e}}_{x'}} + {\alpha _{s,y'}}{E_{y'}}{{\bf{e}}_{y'}} + {\alpha _{s,z'}}{E_{z'}}{{\bf{e}}_{z'}},
\end{equation}
where index $x'$,$y'$, and $z'$ denotes the directions of the three geometric principal axes of the ellipsoidal particle. The polarizability associated with the particle shape is given by \cite{TrojekOptical2012}
\begin{equation}
	{\alpha _{s,i}} = \frac{{\alpha _{s,i}^0}}{{1 - \frac{{i{k^3}\alpha _{s,i}^0}}{{6\pi {\varepsilon _0}}}}},
\end{equation}
\begin{equation}
	\alpha _{s,i}^0 = \frac{{{\varepsilon _0}V}}{{{L_i} + \frac{1}{{{n^2} - 1}}}},
\end{equation}
where $\varepsilon_0$ is the vacuum permittivity and $i=x',y',z'$, $V = \frac{{4\pi }}{3}{r_{x'}}{r_{y'}}{r_{z'}}$ is the particle volume, $r_{x'}$, $r_{y'}$, and $r_{z'}$ are the three semi-axes of the ellipsoid, and $n$ is the refractive index of the particle. The parameter $L_i$ represents the corresponding depolarization factor,
\begin{equation}
	{L_i} = \frac{{{r_{x'}}{r_{y'}}{r_{z'}}}}{2}\int_0^\infty  {\frac{1}{{\left( {s + r_i^2} \right)\sqrt {\left( {s + r_{x'}^2} \right)\left( {s + r_{y'}^2} \right)\left( {s + r_{z'}^2} \right)} }}ds}.
\end{equation}
The real part of the polarizability, ${\mathop{\rm Re}\nolimits} \left[ {{\alpha _{s,i}}} \right] \approx \alpha _{s,i}^0$, determines the optical alignment torque, whereas the imaginary part ${\mathop{\rm Im}\nolimits} \left[ {{\alpha _{s,i}}} \right]$ contributes to optically driven rotation \cite{AhnOptically2018}.

\begin{table}
	\caption{Geometrical parameters of the KTP nanocrystals used to calculate the optical torque arising from shape anisotropy in Supplemental Fig. \ref{fig:s3}(a).}
	\centering
	\label{tab:s2} 
	\begin{tabular}{ccccc}
		\hline
		Particle & $r_{x'}$ (nm) & $r_{y'}$ (nm) & $r_{z'}$ (nm) & $r_{x'} / r_{y'}$ \\
		\hline
		P1 & 60 & 50 & 50 & 1.20 \\
		P2 & 56.6 & 51.5 & 51.5 & 1.10 \\
		P3 & 54.9 & 52.3 & 52.3 & 1.05 \\
		\hline
	\end{tabular}
\end{table}

\begin{figure}
	\centering\includegraphics[width=8cm]{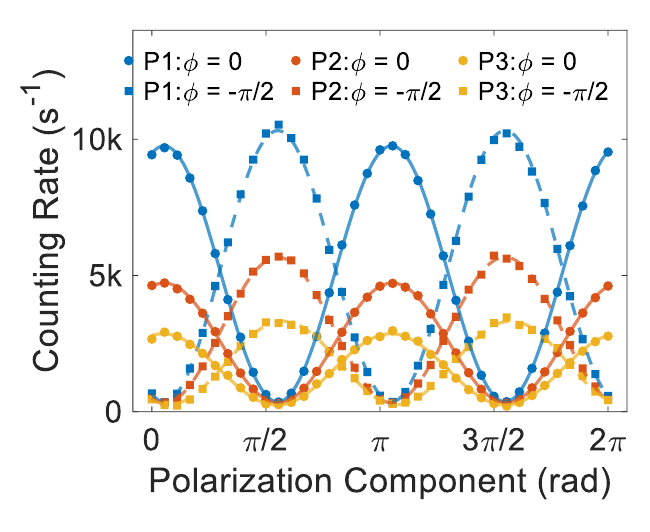}
	\caption{\label{fig:s4} Cartesian-coordinate representation of the polarization-resolved SHG measurements shown in Fig. 4(a) of the main text. Polarization-resolved SHG intensity as a function of analyzer angle for fundamental laser polarization orientations of 0 (circles) and $-\pi/2$ (squares). Measurements from three individual levitated KTP nanocrystals are shown in blue, red, and orange. The solid curves are sinusoidal fits. The identical polarization response observed for particles with different shape confirms that the alignment behavior is dominated by the refractive index anisotropy of KTP rather than by geometric shape effects.}
\end{figure}

To evaluate the shape contribution to the optical torque under the experimental conditions, we consider a trapping laser power of 100 mW and a beam waist of approximately 0.55~$\mu$m at a wavelength of 1064 nm, as measured using the knife-edge method. Since the particle size is much smaller than the beam waist, the electric field can be treated as spatially uniform over the particle volume. For a linearly polarized trapping field along the laboratory $x$ axis, ${\bf{E}} = {{\bf{e}}_x}{E_0}\exp \left( { - i\omega t } \right)$. The polarization direction is assumed to be rotated relative to the particle geometric principal axes by angles $\theta$ and $\phi$, corresponding to rotations about the $y$ and $z$ axes, respectively. The transformation between the laboratory frame $(x,y,z)$ and the particle frame $(x',y',z')$ is described by the rotation matrix
\begin{equation}
	{\bf{R}} = \left( {\begin{array}{*{20}{c}}
			{\cos \phi }&{ - \sin \phi }&0\\
			{\sin \phi }&{\cos \phi }&0\\
			0&0&1
	\end{array}} \right) \cdot \left( {\begin{array}{*{20}{c}}
			{\cos \theta }&0&{\sin \theta }\\
			0&1&0\\
			{ - \sin \theta }&0&{\cos \theta }
	\end{array}} \right) ,
\end{equation}
The electric field in the particle frame is then obtained as $\left( {{E_{x'}},{E_{y'}},{E_{z'}}} \right)$. In the calculations, the refractive index is taken to be $n=1.77$.

The resulting optical torque as a function of the angle between the particle major axis and the polarization direction of the trapping laser is shown in Supplemental Fig. \ref{fig:s3}(a). The calculations are performed for three ellipsoidal particles with identical volumes but different aspect ratios, as shown in Supplemental Table \ref{tab:s2}. The green solid curve corresponds to an ellipsoidal particle with the dimensions of $r_{x'}=60$ nm and $r_{y'}=r_{z'}=50$ nm. The purple dashed and cyan dotted curves are obtained by reducing the aspect ratio of the ellipsoid from 1.20 to 1.10 and 1.05, respectively, while keeping the particle volume constant. As expected, the optical torque decreases as the particle shape approaches a sphere.

\subsection{Optical torque induced by the refractive index anisotropy}

We next consider the optical torque arising solely from the refractive index anisotropy of the KTP crystal. In this case, the induced polarization is given by
\begin{equation}
	{\bf{p}}_r^{\left( 1 \right)} = {\alpha _{r,X}}{E_X}{{\bf{e}}_X} + {\alpha _{r,Y}}{E_Y}{{\bf{e}}_Y} + {\alpha _{r,Z}}{E_Z}{{\bf{e}}_Z},
\end{equation}
where the polarizability associated with crystal birefringence is calculated using the Lorentz–Lorenz relation,
\begin{equation}
	{\alpha _{r,j}} = 3{\varepsilon _0}V\frac{{n_j^2 - 1}}{{n_j^2 + 2}}.
\end{equation}
The indices $j=X,Y,Z$ is the crystal principle axes of KTP. The particle size and the refractive indices used in the calculations are listed in Supplemental Table \ref{tab:s3}.

\begin{table}
	\caption{Geometrical dimensions and principal refractive indices of KTP used for calculating the optical torque arising from refractive index anisotropy in Supplemental Fig. \ref{fig:s4}(b).}
	\centering
	\label{tab:s3} 
	\begin{tabular}{ccccccc}
		\hline
		Particle & $r_{x'}$ (nm) & $r_{y'}$ (nm) & $r_{z'}$ (nm) & $n_X$ & $n_Y$ & $n_Z$ \\
		\hline
		P1 & 60 & 50 & 50 & 1.740 & 1.748 & 1.830 \\
		\hline
	\end{tabular}
\end{table}

Supplemental Fig. \ref{fig:s3}(b) shows the torque arising from the refractive index anisotropy of a KTP sphere as a function of the angle between the crystal $Z$ axis and the polarization direction of the fundamental laser. The green solid and purple dashed curves correspond to rotations involving the $X$ and $Y$ optical axes, respectively.

In experiments, the equilibrium orientation of a levitated KTP nanocrystal is determined by the combined action of shape and refractive index anisotropy. However, for particles with nearly spherical shapes, the torque generated by refractive index anisotropy dominates and becomes the primary mechanism responsible for alignment.

To identify the dominant mechanism experimentally, we randomly characterized three levitated KTP nanocrystals with different shapes. The results are shown in Supplemental Fig. \ref{fig:s4}, where the blue, red, and orange data points correspond to three individual particles, and circles and squares denote horizontal and vertical detection polarizations, respectively. Despite their different geometries, all three particles exhibit nearly identical polarization-alignment behavior. This observation indicates that the orientational dynamics are governed primarily by the refractive index anisotropy of KTP rather than by geometric shape effects.

\end{document}